\documentclass[a4paper,11pt]{article}
\pdfoutput=1 

\usepackage{jcappub} 

\usepackage[T1]{fontenc} 
\usepackage{amsmath,amssymb,amsfonts}
\usepackage{bm}
\usepackage{mathtools}
\usepackage{graphicx}
\usepackage{float}
\usepackage{caption}
\usepackage{subcaption}
\usepackage{booktabs}
\usepackage{array}
\usepackage{multirow}
\usepackage{enumitem}
\usepackage{setspace}
\usepackage{appendix}
\usepackage{geometry}
\usepackage{physics}
\usepackage{mathptmx}
\usepackage{bbm}
\usepackage{threeparttable}
\usepackage[dvipsnames]{xcolor}
\usepackage[numbers,sort&compress]{natbib}
\usepackage{hyperref}
\usepackage{cleveref}

\hypersetup{
    colorlinks=true,
    linkcolor=blue,
    citecolor=blue,
    urlcolor=blue
}
\usepackage{acronym}
\def \OGW {\Omega_{\rm {GW}}}

\newacro{SGWB}[SGWB]{stochastic gravitational-wave background} 
\newacro{FOPT}[FOPT]{first-order phase transition} 
\newacro{SNR}[SNR]{signal-to-noise ratio} 
\newacro{FD}[FD]{flavour deconstruction}
\newacro{SSB}[SSB]{spontaneous symmetry-breaking}
\newacro{LVK}[LVK]{LIGO-Virgo-KAGRA}
\newacro{GW}[GW]{gravitational wave}
\newacro{SM}[SM]{Standard Model}
\newacro{FD}[FD]{Flavour deconstruction}
\newacro{PT}[PT]{phase transition}
\newacro{LISA}[LISA]{Laser Interferometer Space Antenna} 
\newacro{MHD}[MHD]{magnetohydrodynamic} 
\newacro{CI}[CI]{credible interval} 
\newacro{PLS}[PLS]{power-law sensitivity curve}

\title{\boldmath Reconstructing stochastic gravitational-wave signals from flavour deconstruction with LISA}

\author[a,b,1]{N. Fabri,}\note{Corresponding author: \textcolor{blue}{noemi.fabri@physik.uzh.ch}}
\author[b,2]{R. Bertrand-Delgado,}\note{Corresponding author: \textcolor{blue}{raphael.bertrand@physik.uzh.ch}}
\author[b]{D. Laghi,}
\author[a]{L. Mayer}

\affiliation[a]{Institut für Astrophysik, Universität Zürich,\\Winterthurerstrasse 190, CH-8057 Zürich, Switzerland}
\affiliation[b]{Physik-Institut, Universität Zürich,\\ Winterthurerstrasse 190, CH-8057 Zürich, Switzerland}

\emailAdd{noemi.fabri@physik.uzh.ch}
\emailAdd{raphael.bertrand@physik.uzh.ch}
\emailAdd{danny.laghi@physik.uzh.ch}
\emailAdd{lucio.mayer@uzh.ch}

\abstract{
We investigate the reconstruction of a stochastic gravitational-wave background generated by a first-order phase transition in flavour-deconstruction models
using the Laser Interferometer Space Antenna (LISA).
In these scenarios, the spontaneous breaking of an extended flavour-non-universal gauge symmetry at the TeV scale can both generate the observed hierarchies of Standard Model fermion masses and mixing angles and induce a strong first-order phase transition. 
We consider representative benchmark points of the model, compute the corresponding thermodynamic transition parameters, and construct the resulting gravitational-wave spectra using a state-of-the-art sound-wave template. 
We then inject these spectra into simulated LISA data and perform Bayesian inference with \texttt{SGWBinner}, jointly reconstructing the cosmological signal, instrumental noise, and astrophysical foregrounds. 
We find that the strongest benchmark signal can be successfully reconstructed, whereas weaker signals are substantially degenerate with the unresolved extragalactic compact-binary foreground. We further show that tighter prior information on this foreground, motivated by observations with ground-based detectors, can reduce this degeneracy and improve signal reconstruction. 
Our results demonstrate that flavour-deconstruction models can produce signals accessible to LISA, while highlighting the importance of astrophysical-foreground modelling for their identification and characterization.
}

\begin{document}
\maketitle
\flushbottom

\section{Introduction}

A \ac{SGWB} of cosmological origin would provide a direct observational probe of the dynamics of the early Universe and of physics beyond the \ac{SM}~\citep{Christensen:2018iqi, Renzini:2022alw, LIGOScientific:2025kry}.
Among the possible cosmological sources, \acp{FOPT} are particularly well motivated in extensions of the \ac{SM}~\cite{Caprini:2019egz,Caldwell:2022qsj,Caprini:2024hue,Athron:2023xlk}. 
During such a transition, bubbles of the stable phase nucleate and expand within the surrounding plasma. The resulting bulk fluid motion can efficiently source \acp{GW}, whose spectrum retains information about the strength, duration, and temperature of the transition~\citep{Coleman:1977py,Dolan:1973qd,Anderson:1991zb,Hindmarsh:2020hop,Athron:2023xlk}.

In the frequency band relevant to the \ac{LISA}~\cite{LISA:2024hlh}, cosmological backgrounds generated by phase transitions at approximately the electroweak-to-TeV scale are of particular interest~\citep{Caprini:2015zlo,Caprini:2019egz,LISA:2024hlh}.
Their reconstruction will nevertheless be complicated by the presence of instrumental noise and astrophysical foregrounds, including unresolved galactic binaries and extragalactic compact-binary systems~\cite{LISA:2024hlh,Caprini:2024hue}. Assessing the observability of a given particle-physics scenario therefore requires not only predicting its \ac{GW} spectrum, but also determining whether this spectrum can be separated from these additional stochastic components.

\ac{FD} provides a theoretically motivated framework in which the observed flavour hierarchies arise from the spontaneous breaking of an extended, family-dependent gauge symmetry~\citep{Bordone:2017bld,Greljo:2018tuh,Allwicher:2020esa,Isidori:2023sma,Barbieri:2023qpf}.
The associated scalar and gauge sectors can induce strong phase transitions at the TeV scale and generate stochastic \ac{GW} signals potentially accessible to space-based interferometers~\citep{Greljo:2019xan,Fabri:2025FDGW}.
The characteristic frequencies of these signals generally lie above or around the most sensitive part of the \ac{LISA} band, making their observability dependent on both the transition parameters and the ability to disentangle the cosmological signal from astrophysical foregrounds.

In this work, we investigate the reconstruction with \ac{LISA} of the \ac{GW} signals predicted by representative benchmark scenarios of the \ac{FD} model studied in~\cite{Fabri:2025FDGW}. We compute the thermodynamic quantities governing the phase transition from the finite-temperature effective potential and map them onto the corresponding \ac{GW} spectra using the updated sound-wave template of~\cite{Caprini:2024hue}. We then inject these spectra into simulated \ac{LISA} data and perform Bayesian inference with the \texttt{SGWBinner} framework~\cite{Caprini:Reconstructing_spectral_shape_2019,Pieroni:2020rob,Flauger:Improved_recosntruction_2021,Caprini:2024hue}, jointly accounting for the cosmological signal, instrumental noise, and astrophysical foregrounds. Particular attention is devoted to the degeneracy between the phase-transition signal and the unresolved extragalactic compact-binary foreground, and to the extent to which external information on the foreground amplitude can improve the reconstruction.

The paper is organised as follows. Section~\ref{theory} introduces the \ac{FD} framework and describes the computation of the thermodynamic parameters governing the phase transition. Section~\ref{analysis} presents the \ac{GW} signal model and the Bayesian analysis performed with \texttt{SGWBinner}, including the treatment of instrumental noise and astrophysical foregrounds. Section~\ref{discussion} presents and discusses the results. We summarize our conclusions in Section~\ref{conclusions}.

\section{Theoretical framework}
\label{theory}

\subsection{Flavour-deconstruction benchmark}

The \ac{SGWB} considered in this work is generated by a \ac{FOPT} predicted within the \ac{FD} framework~\cite{Houtz:2022fnk, DiLuzio:2018zxy, Fabri:2025FDGW}. We focus on the  4321 realisation, in which the \ac{PT} is associated with the last stage of the spontaneous symmetry-breaking chain \cite{DiLuzio:2018zxy, Greljo:2019xan},
\begin{equation}
     SU(4)^{[3]}\times SU(3)^{[12]} \times U(1)' \longrightarrow
     SU(3)_c \times U(1)_Y\,.
     \label{eq:4321}
\end{equation}
The electroweak group $SU(2)_L$, which is unaffected by this breaking, is left implicit in Eq.~\eqref{eq:4321}. The superscript indicate the fermion families charged under the corresponding flavour-non-universal gauge groups: third-generation quarks and leptons transform under $SU(4)^{[3]}$, whereas the first two generations transform under $SU(3)^{[12]}$.
Matching the extended gauge sector onto the \ac{SM} at the symmetry-breaking scale $v$ imposes
\begin{equation}
g_4^{-2}(v) + g_3^{-2}(v) = g_s^{-2}(v),
\label{eq:matching}
\end{equation}
where $g_4$ and $g_3$ are the couplings of $SU(4)^{[3]}$ and $SU(4)^{[12]}$, respectively, and $g_s(v)$ is the quantum chromodynamics gauge coupling.
Naturalness considerations favour symmetry-breaking scales of a few TeV, with the lowest phenomenologically allowed values being theoretically preferred~\cite{Allwicher:2020esa,Davighi:2023iks}.
Direct searches for the additional gauge bosons 
require approximately
$v\gtrsim1~\mathrm{TeV}$ for $g_4\gtrsim1$ \cite{Haisch:2022afh,Aebischer:2022oqe}. Throughout this work we therefore adopt
\begin{equation}
    v=1~\mathrm{TeV},
\end{equation}
as the reference symmetry-breaking scale.
The spontaneous breaking in Eq.~\eqref{eq:4321} is induced by two complex scalar link fields,
\begin{equation}
\Omega_3\sim(\mathbf{\overline{4}},\mathbf{3},\mathbf{1},\tfrac{1}{6}),
\qquad
\Omega_1\sim(\mathbf{\overline{4}},\mathbf{1},\mathbf{1},-\tfrac{1}{2}),
\end{equation}
where the entries refer to their charge under $SU(4)^{[3]}\times SU(3)^{[12]} \times SU(2)_L \times U(1)'$ respectively.  Following~\cite{Fuentes-Martin:2020hvc,DiLuzio:2018zxy,Fabri:2025FDGW}, we assume that the scalar potential approximately respects a global custodial $SU(4)^{[3]}\times SU(4)^{[12]}$ symmetry, allowing $\Omega_3$ and $\Omega_1$ to be embedded into a single scalar multiplet $\Omega_4$. 
The most general renormalizable scalar potential consistent with this assumption is
\begin{equation}\label{eq:Omega4_Lagr}
\begin{split}
V =&{} \mu^2 \Tr(\Omega_4^\dagger\Omega_4)
+ \rho_1 \left[\Tr (\Omega_4^\dagger\Omega_4)-v^2\right]^2 \nonumber
+ \rho_2 \Tr \left[(\Omega_4^\dagger\Omega_4)-\tfrac12v^2\mathbbm{1}_{4}\right]^2\\
&+ \rho_3 \epsilon_{\alpha \beta \gamma \delta}\epsilon^{\rho \sigma \mu \nu}
(\Omega_4)_\rho^\alpha(\Omega_4)_\sigma^\beta(\Omega_4)_\mu^\gamma(\Omega_4)_\nu^\delta
+ \mathrm{h.c.}.
\end{split}
\end{equation}
In the region of parameter space identified in Ref.~\cite{Fabri:2025FDGW}, 
the radial mode is lighter than the remaining scalar degrees of freedom and its mixing with them is small. The tunneling dynamics can then be approximated by a single real field $\phi$, with tree-level potential
\begin{equation}
V_{\rm tree}(\phi)
=
\frac{\lambda}{4}\phi^4
-\frac12\lambda v^2\phi^2.
\label{eq:V tree}
\end{equation}
For fixed $v$, the dynamics of the transition are therefore controlled primarily by
the effective quartic coupling $\lambda$ and the gauge coupling $g_4$, with $g_3$ determined through Eq.~\eqref{eq:matching}.

\subsection{Finite-temperature effective potential}

The spontaneous symmetry breaking generates fifteen massive gauge degrees of freedom associated with the broken generators.
These are organized into the colour octet $G$, the vector leptoquarks $U$, and the neutral vector boson $Z'$, collectively denoted by $V=\{G,U,Z'\}$. Their field-dependent tree-level masses are
\begin{equation}
m_V(\phi)=g_V\phi,
\end{equation}
and hence $m_V(v) = g_Vv$ in the broken vacuum. The effective couplings are
\begin{equation}
\label{eq:g bosons}
g_G^2=\frac12(g_4^2+g_3^2),\qquad
g_U^2=\frac{1}{2}g_4^2,\qquad
g_{Z'}^2=\frac{1}{2}g_4^2+\frac{1}{3}g_1^2,
\end{equation}
where $g_1$ is the $U(1)'$ coupling.
The corresponding multiplicities are
\begin{equation}
\label{eq:c bosons}
c_G=24,\qquad
c_U=18,\qquad
c_{Z'}=3.
\end{equation}

Following Ref.~\cite{Fabri:2025FDGW}, we retain the dominant gauge-boson contributions to the one-loop effective potential and neglect the subleading fermionic contributions. The finite-temperature contribution is~\cite{Anderson:1991zb}
\begin{equation}
\label{eq:V thermal}
V_{\rm th}(\phi,T)
=
\sum_V
c_VT^4
J_B\!\left(g_V^2\frac{\phi^2}{T^2}\right),
\end{equation}
where, in the convention used here,
\begin{equation}
    J_B(y^2) = \frac{\pi^2}{90} + \frac{1}{2\pi^2} \int_0^{\infty} dx\,x^2 \ln\left(1-e^{-\sqrt{x^2+y^2}}\right).
\end{equation}
The zero-temperature one-loop correction is described by the Coleman--Weinberg potential~\cite{Coleman:1973jx},
\begin{equation}
\label{eq:V CW}
V_{\rm CW}(\phi)
=
\frac1{64\pi^2}
\sum_V
c_Vg_V^4\phi^4
\left(
\ln\frac{g_V^2\phi^2}{\Lambda^2}
-\frac56
\right),
\end{equation}
with renormalization scale $\Lambda=v$. The total effective potential is therefore
\begin{equation}
V_{\rm tot}(\phi,T)
=
V_{\rm tree}
+
V_{\rm CW}
+
V_{\rm th}.
\end{equation}

This calculation is performed at one-loop order. In particular, we do not include daisy resummation or propagate theoretical uncertainties associated with higher-order thermal corrections~\cite{Fabri:2025FDGW}. These effects may modify the location and height of the potential barrier and should be regarded as an uncertainty in the predicted transition parameters.

\subsection{Transition parameters and benchmark selection}
\label{sec:thermodynamics_gw}

The transition is characterized by the nucleation temperature $T_n$, the strength parameter $\alpha$, and the inverse duration 
$\beta/H_n$, where $H_n\equiv H(T_n)$~\cite{Caprini:2015zlo}. We adopt the definitions used in Ref.~\cite{Fabri:2025FDGW}:
\begin{align}
\label{eq:param_Tn} 
T_n &: \qquad \Gamma(T_n) \simeq T_n^4 e^{-S(T_n)} = H_n^4, \\ 
\label{eq:param_alpha} \alpha &: \qquad \alpha = \left. \frac{ V_{\rm tot}(0,T) - V_{\rm tot}(\phi_{\rm min}(T),T) }{ \rho_r(T) } \right|_{T=T_n}, \\ 
\label{eq:param_beta} \frac{\beta}{H_n} &= \left. T\frac{dS(T)}{dT} \right|_{T=T_n}. 
\end{align}
Here, $S(T)\equiv S_3(T)/T$ denotes the finite-temperature Euclidean bounce action, which we compute using \texttt{AnyBubble}~\cite{Masoumi:2016wot}.
The radiation energy density and Hubble rate are
\begin{equation}
\rho_r(T)=\frac{\pi^2}{30}g_*(T)T^4,
\qquad
H^2(T)\simeq\frac{\rho_r(T)}{3M_P^2},
\end{equation}
where $M_P$ is the reduced Planck mass. Following Ref.~\cite{Fabri:2025FDGW}, we set $g_*(T_n)=200$. 

The nucleation condition in Eq.~\eqref{eq:param_Tn} corresponds approximately to the production of one critical bubble per Hubble volume and Hubble time. The parameter $\alpha$ 
quantifies the transition strength relative to the radiation background, while $\beta/H_n$ measures its inverse duration in Hubble units. 
Smaller values of $\beta/H_n$ correspond to a longer transition and generally lead to a larger characteristic bubble separation.

\begin{figure}[t]
    \centering
    \includegraphics[width=1\linewidth]{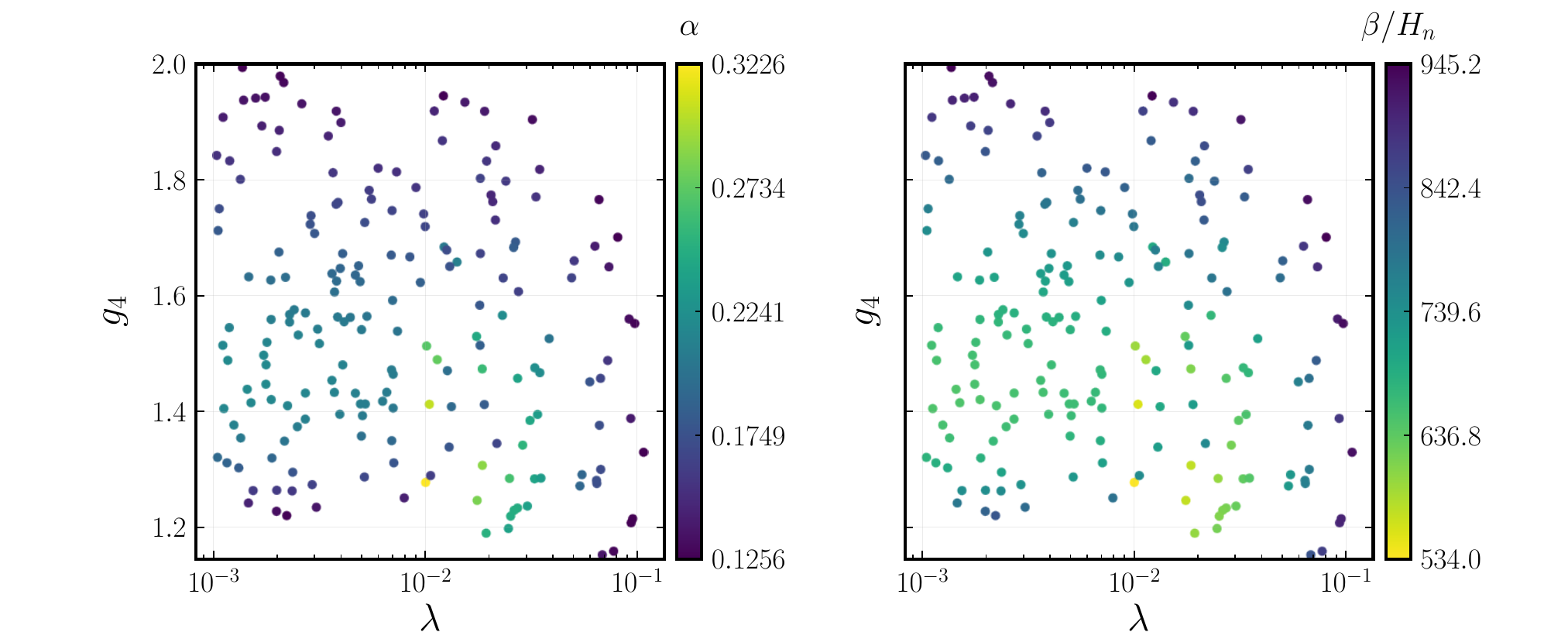}
    \caption{Thermodynamic properties of the \ac{PT} in the $(g_4, \lambda)$ plane
    across the parameter space of the \ac{FD} model for $v=1~{\rm TeV}$. The left panel shows the transition-strength parameter $\alpha$, while the right panel shows the inverse duration $\beta/H_n$. Smaller values of $\lambda$ and gauge couplings around $g_4\sim 1.5$ lead to stronger and longer-lasting transitions. 
    }
    \label{fig:param_space_sgwbinner}
\end{figure}

For the transitions considered here, 
the dominant \ac{GW} source is the acoustic phase generated in the plasma after bubble expansion. 
Contributions from \ac{MHD} turbulence are not expected to contribute over the relevant parameter range and are therefore not included~\cite{Caprini:2024ofd, Caprini:2024hue, Romero:2021kby}.
For $v=1~\mathrm{TeV}$, the parameter-space scan yields
\begin{equation}\label{eq:PT_ranges}
\begin{split}
0.009 &\lesssim \alpha \lesssim 0.3,\\
530 &\lesssim \frac{\beta}{H_n} \lesssim 4500,\\
220 &\lesssim T_n/{\rm GeV} \lesssim 560.
\end{split}
\end{equation}
The parameter scan is performed over the couplings of the flavour-deconstruction model. For each parameter point, the finite-temperature effective potential is evaluated and the Euclidean bounce action is computed numerically, allowing the extraction of the thermodynamic parameters characterizing the phase transition. Only parameter points satisfying the theoretical consistency conditions of the model are retained.
The strongest transitions occur for relatively small values of $\lambda$ and for $g_4 \sim 1.5$, close to the lower boundary implied by the gauge-coupling matching condition in Eq.~\eqref{eq:matching}. Figure~\ref{fig:param_space_sgwbinner} shows the variation of $\alpha$ and $\beta/H_n$ across the scanned $(g_4, \lambda)$ parameter space, showing only the region leading to sizeable first-order phase transitions. As a result, the displayed values of $\beta/H_n$ do not cover the full range reported in Eq.~\eqref{eq:PT_ranges}, since parameter points with very small $\alpha$ and correspondingly large $\beta/H_n$ have been excluded.

We select three representative benchmark points from this region for the reconstruction analysis.
Their thermodynamic parameters are reported in Table~\ref{tab:cases_param_values}. The mapping from these quantities to the corresponding present-day \ac{SGWB} spectra is described in Section~\ref{sec:fopt_model}.

\begin{table}[t]
    \centering
    \renewcommand{\arraystretch}{1.2}
    \begin{tabular}{c c c c}
        \toprule
        Case & $\alpha$ & $\beta/H_n$ & $T_n~[\mathrm{GeV}]$ \\
        \midrule
        1 & $0.323$ & $534$ & $218$ \\
        2 & $0.275$ & $557$ & $225$ \\
        3 & $0.246$ & $618$ & $232$ \\
        \bottomrule
    \end{tabular}
    \caption{Thermodynamic parameters adopted for the three simulated benchmark cases.}
    \label{tab:cases_param_values}
\end{table}

The \ac{GW} template introduced in Section~\ref{analysis} is expressed in terms of quantities evaluated at the time of the \ac{GW} production, denoted by a star~\cite{Caprini:2024ofd}. In the present analysis, we approximate
\begin{equation}\label{eq:Tstar_Tn}
    T_* = T_n, \qquad H_* = H_n.
\end{equation}
This approximation is motivated by the relatively rapid, non-supercooled transitions that we get from the \ac{FD} model. We do not independently compute the percolation temperature $T_p$ \cite{Caprini:2024ofd}, however, and the possible difference between $T_n$ and $T_p$ constitutes an additional theoretical uncertainty.
A further source of theoretical uncertainty is the bubble-wall velocity $\xi_w$, whose determination requires computing the friction exerted by the thermal plasma on the expanding wall \cite{DeCurtis:2022hlx,Laurent:2022jrs,Ai:2023see,DeCurtis:2024hvh,Krajewski:2024gma,Ai:2025bjw,Carena:2025flp,Ekstedt:2024fyq}. Since a dedicated calculation of the wall-plasma dynamics lies beyond the scope of this work, following~\cite{Fabri:2025FDGW}  we adopt the relativistic benchmark $\xi_w=1$.

\section{Methodology}
\label{analysis}

We use the template-based implementation of 
\texttt{SGWBinner} to simulate and reconstruct the stochastic components of the \ac{LISA} data~\citep{Caprini:Reconstructing_spectral_shape_2019,Flauger:Improved_recosntruction_2021,Caprini:2024hue}.

\subsection{Sound-wave signal model}
\label{sec:fopt_model}

For the transitions considered here, we model the \ac{SGWB} using the sound-wave template implemented in \texttt{SGWBinner}~\citep{Kamionkowski:1993fg, Kosowsky:1992rz, Huber:2008hg, Lewicki:2022pdb, Hindmarsh:2013xza, Hindmarsh:2015qta, Hindmarsh:2017gnf, Hindmarsh:2019phv,Caprini:2024hue}. Our baseline assumption is therefore
\begin{equation} 
h^2\Omega_{\rm FOPT}(f) \simeq h^2\Omega_{\rm sw}(f). 
\label{eq:sound_only} 
\end{equation}
We neglect a possible contribution from magnetohydrodynamic turbulence~\cite{Cutting:2019zws, Correia:2025qif, Caprini:2015zlo}. The estimate supporting this approximation is presented in Appendix~\ref{sec:turbulent_motion}. This should be nevertheless regarded as a modelling assumption rather than a direct consequence of the value of $\alpha$, particularly because the acoustic lifetime of the benchmark transitions is found to be shorter than one Hubble time \citep{Caprini:2019egz, Ellis:lifetime_sound_waves_2020}.

The sound-wave spectrum is described in terms of the thermodynamic parameters
\begin{equation}\label{eq:thermo_params}
\boldsymbol{\theta}_{\rm PT} = \{K,H_*R_*,\xi_w,T_*\}, 
\end{equation}
where $K$ is the fraction of the total energy density converted into bulk kinetic energy, $R_*$ is the characteristic bubble separation, $\xi_w$ is the bubble-wall velocity, and $T_*$ is the temperature at the time of \ac{GW} production.

For a relativistic wall, we calculate the kinetic-energy fraction as
\citep{espinosaEnergyBudgetCosmological2010,Caprini:2015zlo,Caprini:2019egz,Kamionkowski:1993fg,Huber:2008hg,Lewicki:2022pdb}:
\begin{align}
    \label{eq:K} K &= \kappa_{sw}\frac{\alpha}{1+\alpha} \ ,\\
    \label{eq:kappa}
    \kappa_{\rm sw} &= \frac{\alpha}{0.73+0.083\sqrt{\alpha}+\alpha} \ .    
\end{align}
We remark that Eq.~\eqref{eq:K} follows a different prescription than that in~\cite{Caprini:2024hue}, which includes an additional factor of $0.6$, accounting for the reduced efficiency of kinetic energy production relative to an isolated expanding bubble. 

The characteristic bubble separation is related to the inverse duration of the transition by~\citep{Caprini:2019egz, Caprini:2024hue}
\begin{equation}
    \label{eq:HR_star} H_* R_* = (8\pi)^{1/3} {\rm max}(\xi_w, c_s) \frac{H_*}{\beta},
\end{equation}
where $c_s=1/\sqrt{3}$ is the sound speed in a radiation fluid. For the injected spectra, we adopt the relativistic-wall approximation $\xi_w = 1$~\citep{Fabri:2025FDGW}, so that 
\begin{equation}
    H_*R_* = (8\pi)^{1/3}\frac{H_*}{\beta}.
\end{equation}
The wall velocity $\xi_w$ is nevertheless allowed to vary in the inference.

The logarithmically integrated sound-wave amplitude is modelled as~\cite{Caprini:2024hue}:
\begin{equation}
    \label{eq:omega_sw_1} 
    h^2 \Omega_{\rm sw} = h^2 A_{\rm sw} \left(\frac{a_*}{a_0}\right)^4 \left(\frac{H_*}{H_0}\right)^2 K^2 (H_* \tau_{\rm sw}) ( H_*R_*),
\end{equation}
where $A_{\rm sw} = 0.11$~\citep{Jinno:2022mie, Caprini:2024gyk, Caprini:2024hue} and
\begin{equation} 
H_*\tau_{\rm sw} = \min\left( 1, \frac{H_*R_*}{\bar v_f} \right), \qquad \bar v_f = \sqrt{\frac{K}{\Gamma}}, \qquad \Gamma=\frac{4}{3}, 
\label{eq:sound_lifetime}
\end{equation}
where $\Bar{v}_f $ is the average fluid velocity and $\Gamma$ the mean adiabatic index equal to $4/3$ for radiation fluids \citep{Caprini:2019egz}.
For the benchmark points considered here,
\begin{equation} 
\frac{H_*R_*}{\bar v_f} \sim10^{-2}, 
\end{equation}
and the signal is therefore in the short-lived acoustic regime,
\begin{equation}
     H_* \tau_{\rm sw} = \frac{H_*R_*}{\Bar{v}_f}. 
\end{equation}
Eq.~\eqref{eq:omega_sw_1} then becomes
\begin{equation}
\label{eq:sw_amplitude_int}
    h^2 \Omega_{\rm sw} = h^2 A_{\rm sw} \left(\frac{a_*}{a_0}\right)^4 \left(\frac{H_*}{H_0}\right)^2 \frac{K^2  ( H_*R_*)^2}{\Bar{v}_f} \  .
\end{equation}

The spectral shape is described by a double broken power law. We write it in the parametrization used internally by \texttt{SGWBinner},
\begin{equation} \label{eq:omega_dbpl_reparam} h^2\Omega_{\rm sw}(f) =  h^2\Omega_2 S_2(f;f_1,f_2), \end{equation}
with  $\int^{\infty}_{-\infty} \dd \ln f h^2\Omega_{\rm sw}(f) = h^2\Omega_{\rm sw}$ defined in Eq. \eqref{eq:sw_amplitude_int} and
\begin{equation} 
S_2(f;f_1,f_2)  = \mathcal{N}_2 \left(\frac{f}{f_1}\right)^3 \left[ 1+\left(\frac{f}{f_1}\right)^2 \right]^{-1} \left[ 1+\left(\frac{f}{f_2}\right)^4 \right]^{-1} , 
\label{eq:dbpl_shape} 
\end{equation}
where $\mathcal{N}_2$ is fixed such that $S_2(f_2) = 1$. Accordingly, $\Omega_2\equiv\Omega_{\rm sw}(f_2)$ is the amplitude at the second frequency break. This should not be identified in general with the spectral peak or with the logarithmically integrated amplitude~\cite{Caprini:2015zlo,Caprini:2024hue}.

The corresponding geometric parameters are
\begin{equation}\label{eq:geom_params}
    \boldsymbol\theta_{\rm geo} = \{\Omega_2, f_1, f_2  \}.
\end{equation}
The two break frequencies are related to the thermodynamic parameters through~\cite{Caprini:2015zlo, Caprini:2024hue} 
\begin{equation} 
\label{eq:freqency_breaks}
    f_1 \simeq 0.2 \, H_{*,0}(H_*R_*)^{-1} , \qquad f_2\simeq 0.5 \, H_{*,0} \Delta_w^{-1}(H_*R_*)^{-1},
\end{equation}
where 
\begin{equation} 
H_{*,0} = \frac{a_*}{a_0}H_*, \qquad \Delta_w = \frac{|\xi_w-c_s|} {\max(\xi_w,c_s)}. 
\end{equation}
For the relativistic-wall injection, 
$\xi_w = 1 > c_s$ 
and hence $\Delta_w = 1 - c_s$.

The mapping from the four thermodynamic parameters defined in Eq.~\eqref{eq:thermo_params} to the three geometric parameters defined in Eq.~\eqref{eq:geom_params} is not one-to-one. Significant correlations and degeneracies are therefore expected when the thermodynamic parameters are reconstructed directly~\cite{Caprini:2024hue}. We adopt independent uniform priors on these parameters, with the ranges reported in Table~\ref{tab:fopt_priors}.

\begin{figure}[t]
     \centering
     \includegraphics[width=\linewidth]{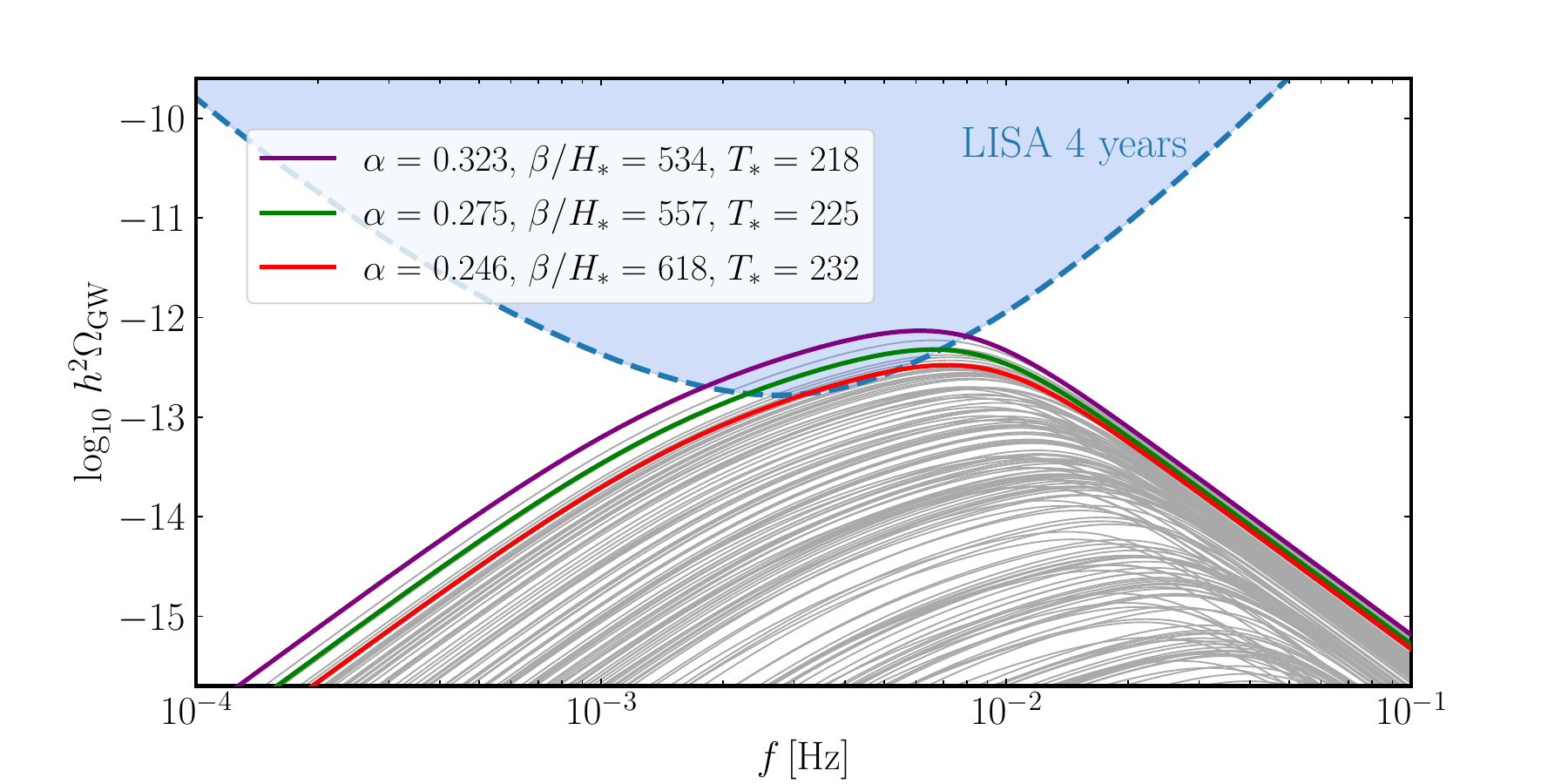}
     \caption{Representative stochastic \ac{GW} spectra generated by the \ac{FD} model for $v=1~{\rm TeV}$. The coloured curves show the three benchmark spectra analysed in this work, while the grey curves represent additional points from the parameter-space scan shown in Figure~\ref{fig:param_space_sgwbinner}. The dotted blue curve shows the four-year power-law integrated sensitivity of \ac{LISA}. This curve provides a useful visual reference but does not account for the simultaneous reconstruction of astrophysical foregrounds.}
     \label{fig:gw-parm-change}
\end{figure}

Importantly, the geometric parameters are deterministic functions of the thermodynamic parameters and are therefore not assigned independent priors. Applying the mapping described above to the three benchmark transitions in Table~\ref{tab:cases_param_values} gives the present-day spectra shown in Fig.~\ref{fig:gw-parm-change}, which also shows other curves obtained using the points shown in Fig.~\ref{fig:param_space_sgwbinner}. 
The signal amplitude increases with the kinetic-energy fraction $K$ and with the characteristic bubble separation $H_*R_*$. 
The transition temperature primarily controls the redshifted frequency scale, while the wall velocity affects both the characteristic bubble separation and the separation between the two spectral breaks. 
For comparison with conventional sensitivity forecasts, Fig.~\ref{fig:gw-parm-change} also shows the four-year power-law-integrated sensitivity of \ac{LISA}, calculated for the same observation time and instrumental-noise model adopted in the analysis. 
Its construction is described in Appendix~\ref{app:pli}. 
Since the phase-transition spectra are not pure power laws and the curve does not account for the simultaneous reconstruction of astrophysical foregrounds, it is included only as a visual sensitivity reference.

\begin{table}[t]
    \centering
    {\renewcommand{\arraystretch}{1.2}
    \begin{tabular}{c c}
        \hline
        Parameter & Prior range \\
        \hline
        $\log_{10}(K)$
            & $[-4,-0.2]$ \\
        $\log_{10}(H_*R_*)$
            & $[-5,0]$ \\
        $\xi_w$
            & $[0.6,1]$ \\
        $\log_{10}(T_*/{\rm GeV})$
            & $[0,6]$ \\
        \hline
    \end{tabular}}
    \caption{Uniform prior ranges adopted for the \ac{FOPT} parameters.}
    \label{tab:fopt_priors}
\end{table}

\subsection{Astrophysical foregrounds and instrumental noise}
\label{sec:foregrounds_noise}

In addition to the \ac{FOPT} signal, 
we include two astrophysical foregrounds in both the simulated data and the reconstruction. 
The extragalactic foreground is generated by the superposition of unresolved stellar-mass compact binaries~\citep{Zhu:2011bd, Renzini:2022alw, Meacher:2014aca, LIGOScientific:2025bgj, Babak:2023lro, Bavera:2021wmw}. In the \ac{LISA} band, these systems are predominantly observed during their inspiral and their combined signal can be approximated by~\citep{Phinney:2001di, Regimbau:2011rp}:
\begin{equation}
    h^2\Omega_{\rm GW, Ext}(f) = h^2 \Omega_{\rm Ext} \left(\frac{f}{f_{\rm Ext,ref}} \right)^{2/3}, \qquad f_{\rm Ext,ref}=1~\mathrm{mHz}.
\end{equation}

The unresolved galactic compact-binary foreground is modelled as~\citep{Karnesis:2021tsh,Robson:2018ifk}: 
\begin{equation} \label{eq:fg_galactic}
    h^2 \Omega_{\rm GW, Gal}(f) = h^2 \Omega_{\rm Gal} \frac{f^3}{2}\left(\frac{f}{f_{\rm Gal, ref}} \right)^{-7/3} e^{-(f/f_{1,\rm Gal})^v}\left[ 1+ \tanh \left( \frac{f_{\rm knee} - f}{f_{2,\rm Gal}}\right)\right]   \ , 
\end{equation}
where $f_{\rm Gal, ref} = 1~\mathrm{Hz}$. The exponential suppression describes the loss of stochasticity at high frequencies, while the hyperbolic-tangent factor models the progressive resolution and subtraction of individual systems.
We use the four-year fiducial values $f_{1,\rm Gal}$, $f_{2,\rm Gal}$, $f_{\rm knee}$, and $\nu$ given in Ref.~\cite{Karnesis:2021tsh}, and vary only the overall amplitude $\Omega_{\rm Gal}$.

The total stochastic signal is therefore
\begin{equation}
    \OGW (f)= \Omega_{\rm FOPT}(f) +  \Omega_{\rm GW, Ext}(f) + \Omega_{\rm GW, Gal}(f) \ ,
\end{equation}

When neither break of the \ac{FOPT} spectrum is well measured in the \ac{LISA} band, the observable part of the signal can resemble a power law. It may then become strongly degenerate with the extragalactic compact-binary foreground~\citep{Kume:Assessing_impact_2025, Boileau:Prospects_FOPT_detection_2022}.
To assess the effect of external information on this degeneracy, we perform analyses using both a broad prior and an informative prior on $\Omega_{\rm Ext}$. The latter has a standard deviation of $0.02$ in $\log_{10}(h^2\Omega_{\rm Ext})$, corresponding approximately to a $5\%$ relative uncertainty. We treat this as an illustrative externally informed prior motivated by the expected improvement in compact-binary population constraints~\citep{Babak:2023lro}.

For the instrumental noise, we assume an equilateral \ac{LISA} configuration with identical noise levels in the three arms. The dominant contributions arise from the test masses and the optical measurement system~\citep{Tinto:2004wu, Vallisneri:2005ji, Hartwig:2021mzw,Caprini:2024hue, Babak:2021mhe}: 
\begin{align}
    \label{eq:noise_TM} S^{\rm TM}(f) &= A^2 \left[ 1+ \left(\frac{0.4\ {\rm mHz}}{f}\right)^2\right] \left[ 1+ \left(\frac{f}{8\ {\rm mHz}}\right)^4\right] \left(\frac{1}{2\pi f c}\right)^2 \frac{{\rm fm}^2}{{\rm s}^3} \ ,\\
    \label{eq:noise_OMS} S^{\rm OMS}(f) &= P^2 \left[ 1+ \left(\frac{2\, {\rm mHz}}{f}\right)^4\right] \left(\frac{2\pi f}{ c}\right)^2 \frac{{\rm pm}^2}{{\rm Hz}} \ .    
\end{align}
The amplitudes $A$ and $P$ are treated as free parameters, while $c$ is the speed of light. Gaussian priors are adopted for the foreground and instrumental-noise parameters, as summarized in Table~\ref{tab:foreground_noise_priors}.

\begin{table}[t]
    \centering
    {\renewcommand{\arraystretch}{1.3}
    \begin{tabular}{c c c | c c c}
        \hline
        \multicolumn{3}{c|}{Astrophysical foregrounds}
        & \multicolumn{3}{c}{Instrumental noise} \\
        Parameter & $\mu$ & $\sigma$
        & Parameter & $\mu$ & $\sigma$ \\
        \hline
        $\log_{10}\!\left(h^2\Omega_{\rm Ext}\right)$
        & $-12.38$
        & $0.17\;(0.02)$
        & $A$
        & $3$
        & $0.6$ \\

        $\log_{10}\!\left(h^2\Omega_{\rm Gal}\right)$
        & $-7.84$
        & $0.21$
        & $P$
        & $15$
        & $3$ \\
        \hline
    \end{tabular}}
    \caption{Means, $\mu$, and standard deviations, $\sigma$, of the Gaussian priors adopted for the astrophysical-foreground and instrumental-noise parameters. The value in parentheses corresponds to the narrower prior on the extragalactic-foreground amplitude.}
    \label{tab:foreground_noise_priors}
\end{table}

\subsection{Simulated LISA data and likelihood construction}
\label{sec:likelihood}

We analyse the orthogonal time-delay interferometry  channels $I\in\{A,E,T\}$~\cite{Caprini:2024hue}. Under the assumption of an equilateral constellation and identical arm noise, these channels have negligible mutual correlations. The $A$ and $E$ channels carry most of the \ac{GW} sensitivity, while the approximately null $T$ channel helps constrain the instrumental noise.

In the frequency domain, the stochastic data in channel $I$ are written as
\begin{equation}
    \widetilde d_I(f) = \sum_\nu\widetilde n_I^\nu(f) + \sum_\sigma\widetilde s_I^\sigma(f),
\end{equation}
where $\nu$ labels the instrumental-noise components and $\sigma$ labels the cosmological and astrophysical signals. We assume these components to be independent, stationary, zero-mean Gaussian random variables. Their two-point function is
\begin{equation} 
\left\langle \widetilde d_I(f) \widetilde d_J^*(f') \right\rangle = \frac{1}{2} \delta_{IJ} \delta^{\rm D}(f-f') \left[ P_{n,II}(f)+P_{s,II}(f) \right], 
\end{equation}
where $\delta_{IJ}$ and $\delta^D(f- f')$ are Kronecker and Dirac deltas, respectively, while $ P_{n,II}(f)$ and $P_{s,II}(f)$ are respectively the noise and signal channel power spectra, which 
include the corresponding \ac{LISA} response functions, which are fixed to the values implemented in \texttt{SGWBinner}~\cite{Caprini:2024hue, Caprini:Reconstructing_spectral_shape_2019}.

We divide the data into $N_d = 126$ segments of duration $\tau = 11.4$ days, summing up to $T_{\rm obs}$ = 4 effective years of observation~\cite{LISA:2024hlh,Caprini:2024hue, Caprini:Reconstructing_spectral_shape_2019}. For each frequency sample $f_k$, the segment-averaged auto-power estimator is
\begin{equation} 
\overline D_{II}(f_k) = \frac{1}{N_d} \sum_{s=1}^{N_d} \widetilde d_{I,s}(f_k) \widetilde d_{I,s}^*(f_k). 
\label{eq:segment_average} 
\end{equation} 
The initial high-resolution frequency grid spans 
\begin{equation} f_{\min}=3\times10^{-5}~\mathrm{Hz}, \qquad f_{\max}=5\times10^{-1}~\mathrm{Hz}. 
\end{equation}
For computational efficiency, \texttt{SGWBinner} subsequently coarse-grains these data using inverse-variance weighting~\cite{Caprini:2024hue}. We denote the resulting data by $D_{II}^k$, where $k$ labels the coarse-grained frequency bins, and by $w_{II}^k$ the corresponding statistical weights.

The likelihood combines Gaussian and log-normal contributions~\citep{Caprini:2024hue},
\begin{equation}
\label{eq:combined_likelihood} 
\ln\mathcal{L}(\boldsymbol{\theta}) = \frac{1}{3} \ln\mathcal{L}_{\rm G}(\boldsymbol{\theta}) + \frac{2}{3} \ln\mathcal{L}_{\rm LN}(\boldsymbol{\theta}), 
\end{equation}
where $\boldsymbol{\theta}$ is the parameter vector to be inferred and  
\begin{equation} 
\label{eq:gaussian_likelihood} 
\ln\mathcal{L}_{\rm G} = -\frac{N_d}{2} \sum_k\sum_I w_{II}^k \left[ 1- \frac{D_{II}^k} {D_{II}^{\rm Th}(f_k;\boldsymbol{\theta})} \right]^2, 
\end{equation} 
and 
\begin{equation} 
\label{eq:lognormal_likelihood} 
\ln\mathcal{L}_{\rm LN} = -\frac{N_d}{2} \sum_k\sum_I w_{II}^k \left[ \ln \frac{ D_{II}^{\rm Th}(f_k;\boldsymbol{\theta}) }{ D_{II}^k } \right]^2. 
\end{equation} 
Here, $D_{II}^{\rm Th}$ is the theoretical prediction including the \ac{FOPT} signal, both astrophysical foregrounds, and instrumental noise. The log-normal term accounts approximately for the mild non-Gaussianity introduced by the coarse-graining procedure~\cite{Caprini:2024hue}.

\subsection{Bayesian inference and summary statistics}

We sample the posterior distribution using \texttt{PolyChord}~\cite{Handley:2015vkr, Handley:2015fda}. 
For a model with $d$ free parameters, we use
\begin{equation} 
n_{\rm live}=200d, \qquad n_{\rm rep}=20d, \qquad \epsilon_{\rm precision}=10^{-6}. 
\end{equation}
with $d=8$ in all our analyses.
Unless stated otherwise, inferred parameters are summarized by their posterior medians and 68\% \ac{CI}.

For comparison with conventional sensitivity forecasts, we calculate the optimal \ac{SNR} of the injected \ac{FOPT} signal, 
\begin{equation} 
\label{eq:SNR} 
\rho_{\rm inj}^2 = T_{\rm obs} \sum_{I=A,E,T} \int_{f_{\min}}^{f_{\max}} df\, \left[ \frac{ h^2\Omega_{{\rm FOPT},I}^{\rm inj}(f) }{ h^2\Omega_{N,I}(f) } \right]^2. 
\end{equation}
Here, $h^2\Omega_{{\rm FOPT},I}^{\rm inj}(f)$ is the injected phase-transition spectrum projected onto TDI channel $I$, while $h^2\Omega_{N,I}(f)$ is the corresponding equivalent instrumental-noise spectrum including the combined test-mass and optical-measurement-system contributions~\cite{Caprini:2024hue}. 
Both quantities include the appropriate \ac{LISA} response function. 

The injected SNR depends only on the specified signal, instrumental noise, observation time, and analysis band. It is therefore independent of the priors adopted in the Bayesian reconstruction. 
We also define a reconstructed SNR by evaluating the same expression for the phase-transition spectrum associated with each posterior sample, 
\begin{equation} 
\label{eq:SNR_rec} 
\left[\rho_{\rm rec}^{(r)}\right]^2 = T_{\rm obs} \sum_{I\in\{A,E,T\}} \int_{f_{\min}}^{f_{\max}} df\, \left[ \frac{ h^2\Omega_{{\rm FOPT},I} \bigl(f;\boldsymbol{\theta}_{\rm PT}^{(r)}\bigr) }{ h^2\Omega_{N,I}(f) } \right]^2, 
\end{equation} 
where $r$ labels a posterior sample. We evaluate Eq.~\eqref{eq:SNR_rec} for every posterior sample and report the median and 68\% confidence interval of the distribution. 
First, we translate the posteriors of thermodynamic parameters into the geometrical ones to obtain a posterior distribution of the estimated signal. Then, we compute the SNR distribution from the geometrical posteriors.

Both $\rho_{\rm inj}$ and $\rho_{\rm rec}$ are used only as summary diagnostics. A high reconstructed SNR does not by itself establish that the data favour a model containing a \ac{FOPT} component over a foreground-and-noise-only model~\cite{Caprini:2024hue}. Such a detection claim would require an explicit model comparison, which is not performed here.

The power-law-integrated sensitivity shown in Fig.~\ref{fig:gw-parm-change} is constructed from the same instrumental-noise spectra and observation time, adopting a threshold $\rho_{\rm th}=10$~\cite{Caprini:2015zlo, Bartolo:Science_case_LISA_2016}. The full construction is described in Appendix~\ref{app:pli}. This curve is not used in the Bayesian inference and should not be interpreted as a reconstruction boundary in the presence of astrophysical foregrounds.

As a complementary consistency check, in the strongest signal case we also estimate local parameter uncertainties using a Fisher-information analysis. 
This approach approximates the likelihood by a multivariate Gaussian around the injected parameter values and provides a computationally inexpensive comparison with the nested-sampling results. 
Its construction, including the treatment of foreground and instrumental-noise parameters, is presented in Appendix~\ref{app:fisher}. 
Fisher estimates are used for geometric parameters that should not be degenerate, in comparison to thermodynamic parameters which exhibit strong correlations, non-Gaussian posterior tails, and sensitivity to prior boundaries.
We stress that the Fisher analysis serves only as a diagnostic and not as the primary reconstruction result.

Finally, the simulated data and the reconstruction use the same functional forms for the cosmological signal, astrophysical foregrounds, and instrumental noise. We also neglect subtraction residuals from individually resolved sources and theoretical uncertainties in the phase-transition template. The resulting forecasts should therefore be interpreted as an idealized assessment of parameter reconstruction~\cite{Caprini:2024hue}.

\section{Results and discussion}
\label{discussion}

We analyse the three benchmark \ac{PT} introduced in Table~\ref{tab:cases_param_values} and shown in Figure~\ref{fig:gw-parm-change}.
Their injected \acp{SNR} are $\rho_{\rm inj}=$ 32.3, 19.7, and 12.8, respectively. The corresponding reconstructions results are summarized in Table~\ref{tab:snr_68CL}.
Case 1 represents the strongest benchmark and can be reconstructed using the broad foreground priors of Table~\ref{tab:foreground_noise_priors}. Cases 2 and 3 instead illustrate the increasing degeneracy between the cosmological signal and the unresolved extragalactic compact-binary foreground. For these two cases, we therefore compare the baseline analysis with the analysis employing the narrower extragalactic-foreground prior introduced in Section~\ref{sec:foregrounds_noise}.

\subsection{Case 1: reconstruction of the strongest benchmark}
\label{sec:detectable_signal}

\begin{figure}[t]
    \centering
    \includegraphics[width=1\linewidth]{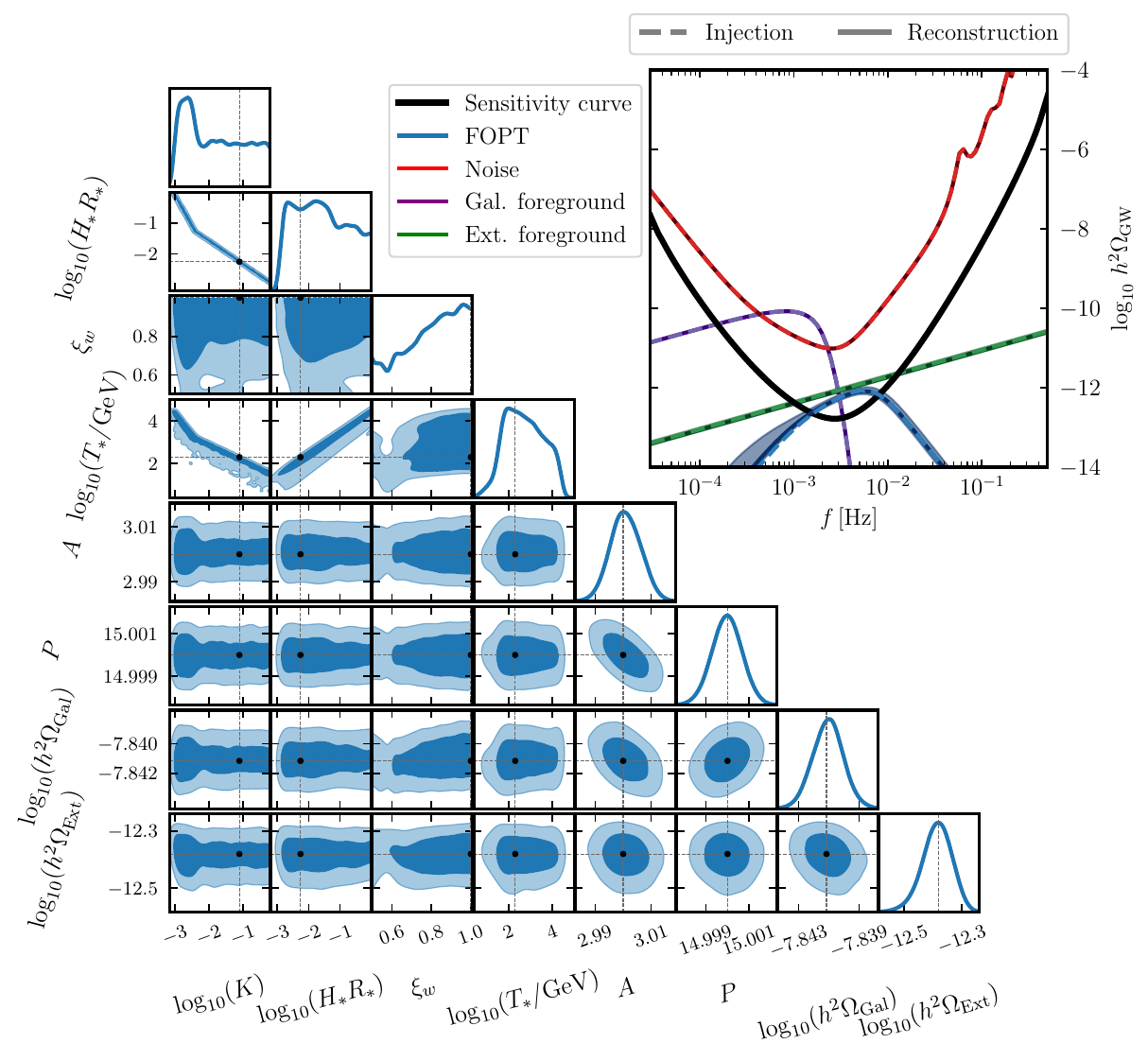}
    \caption{
     Bayesian reconstruction for Case~1, using the broad extragalactic-foreground prior. The corner plot shows the marginalized posterior distributions of the \ac{FOPT}, astrophysical-foreground, and instrumental-noise parameters, with the $68\%$ and $95\%$ credible regions shown in dark and light blue, respectively. Black markers and dashed lines denote the injected values. The upper-right panel shows the reconstructed stochastic components: the \ac{FOPT} signal, astrophysical foregrounds, and instrumental noise. The shaded band denotes the $68\%$ posterior interval of the reconstructed \ac{FOPT} spectrum. The power-law integrated sensitivity is shown for reference only and is not used in the Bayesian reconstruction.
    }
    \label{fig:run_1_corner}
\end{figure}

Figure~\ref{fig:run_1_corner} shows the posterior distributions and spectral reconstructions for Case 1. The injected signal has $\rho_{\rm inj}=32.3$, while the reconstructed-SNR posterior is $\rho_{\rm rec} = 35.8^{+6.9}_{-7.1}$, where the quoted value and uncertainties denote the posterior median and $68\%$ \ac{CI} obtained from Eq.~\eqref{eq:SNR_rec}. The reconstructed-SNR distribution is therefore shifted towards somewhat larger values than the injected SNR. However, this difference should not be interpreted as a detection significance, since $\rho_{\rm rec}$ is a derived reconstruction diagnostic and not a model-comparison statistic.

The cosmological component is nevertheless clearly recovered within the assumed signal model. The instrumental-noise and astrophysical foreground parameters are simultaneously constrained, with the extragalactic foreground recovered as $\log_{10}(h^2\Omega_{\rm Ext}) = -12.383^{+0.054}_{-0.045}$, consistent with its injected value. The instrumental-noise and galactic-foreground parameters are even more tightly constrained and are omitted from Table~\ref{tab:snr_68CL} for clarity. 

Despite the successful spectral reconstruction, the individual thermodynamic parameters remain only weakly constrained. We obtain
\begin{align} 
\log_{10}K &= -1.6^{+1.3}_{-1.1}, &
\log_{10}(H_*R_*) &= -1.88^{+0.59}_{-0.92}, \\
\xi_w &= 0.792^{+0.180}_{-0.087}, & 
\log_{10}(T_*/{\rm GeV}) &= 2.44^{+0.61}_{-0.84}. 
\end{align}
The broad and strongly correlated posteriors are a consequence of the mapping between the four thermodynamic parameters $\boldsymbol{\theta}_{\rm PT}$ and only three independent spectral parameters $\boldsymbol{\theta}_{\rm geo}$.
In particular, during radiation domination, $H_{*,0} = (a_*/a_0)H_* \propto T_* g_*^{1/6}$, so that  Eq.~\eqref{eq:freqency_breaks} implies $f_1 \propto T_*/(H_* R_*)$ for fixed relativistic degrees of freedom. 
A measurement of the first break therefore primarily constrains the ratio $T_*/(H_*R_*)$, leaving the two parameters individually degenerate. 
Similarly, in the short-lived regime relevant for the present benchmarks, Eq.~\eqref{eq:sw_amplitude_int} gives approximately $\Omega_{\rm sw} \propto K^{3/2}(H_* R_*)^2$, so that different combinations of $K$ and $H_* R_*$ can generate similar amplitudes.

\begin{figure}[h]
    \centering
    \includegraphics[width=0.55\linewidth]{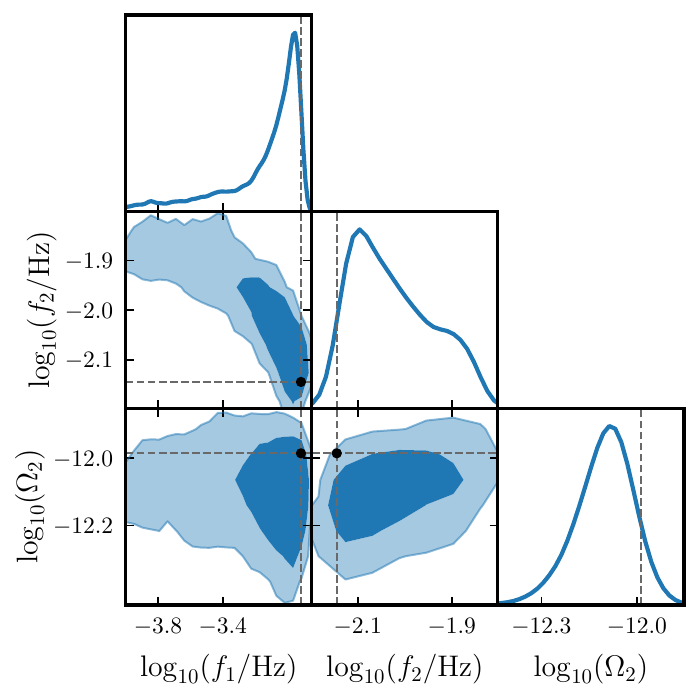}
    \caption{Reconstruction of the 
    derived spectral parameters for Case~1, obtained by mapping the thermodynamic posterior samples through the sound-wave template of Section~\ref{sec:fopt_model}. The $68\%$ and $95\%$ credible regions are shown in dark and light blue, respectively, while the black marker and dashed lines denote the values corresponding to the injected signal.
    }
    \label{fig:run_1_geometric_corner}
\end{figure}

The distinction between spectral and thermodynamic reconstruction is illustrated in Fig.~\ref{fig:run_1_geometric_corner}. Transforming the posterior samples to quantities that directly characterize the double-broken-power-law spectrum through Eqs. \eqref{eq:sw_amplitude_int} and \eqref{eq:freqency_breaks} yields considerably tighter constraints on the amplitude and on the second frequency break. The first break, $f_1$, remains more weakly and asymmetrically constrained, consistently with the relative poor determination of $\xi_w$ and of the low-frequency part of the signal. The corresponding posterior summaries are reported in Table~\ref{tab:post_geom_param}.

\begin{table}[h]
    \centering
    \renewcommand{\arraystretch}{1.2}
    \begin{tabular}{c c c}
        \toprule
        $\log_{10}\!\left(\Omega_2\right)$
        & $\log_{10}\!\left(f_1/\mathrm{Hz}\right)$
        & $\log_{10}\!\left(f_2/\mathrm{Hz}\right)$ \\
        \midrule
        $-12.10^{+0.10}_{-0.08}$
        & $-3.17^{+0.28}_{-0.01}$
        & $-2.03^{+0.06}_{-0.12}$ \\
        \bottomrule
    \end{tabular}
    \caption{
       Posterior median values and corresponding $68\%$ \ac{CI} for the reconstructed spectral parameters for Case~1.}
    \label{tab:post_geom_param}
\end{table}

\begin{figure}[h]
    \centering
    \includegraphics[width=\linewidth]{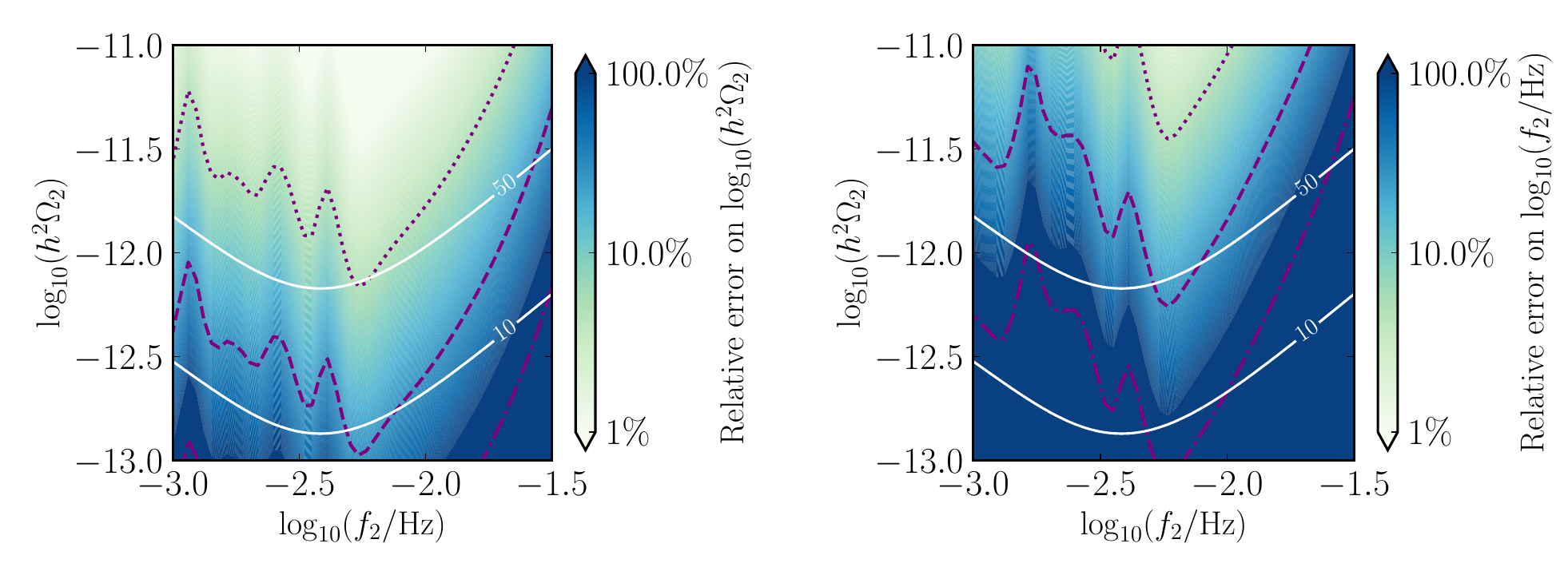}
    \caption{
    Fisher forecast for the relative uncertainty on the amplitude at the second spectral break, $\Omega_2$ (left), and on the second break frequency, $f_2$ (right), for the double broken-power-law template, for Case~1. The purple contours indicate fixed fractional uncertainties: 5\%, 30\%, and 200\% for the dotted, dashed, and dash-dotted lines, respectively. The white contours indicate constant optimal \ac{SNR}. The non-monotonic structure reflects the frequency dependence of the combined instrumental-noise and foreground spectra. The Fisher calculation is used only as a diagnostic comparison with the full Bayesian reconstruction, see Appendix~\ref{app:fisher}.
    }
    \label{fig:relative_error_fisher}
\end{figure}

The same behaviour is qualitatively reproduced by the Fisher analysis shown in Fig.~\ref{fig:relative_error_fisher}. The Fisher calculation is performed directly in terms of spectral quantities and therefore does not suffer from the intrinsic four-to-three mapping degeneracy of the thermodynamic parametrization. Its detailed construction is given in Appendix~\ref{app:fisher}. We use this comparison only as a local consistency check, since the Bayesian posteriors can be non-Gaussian and affected by prior boundaries.

\subsection{Cases 2 and 3: foreground-limited reconstruction}
\label{sec:degenerate_signal}

The weaker benchmarks illustrate a qualitatively different limitation: even when the injected optimal SNR is appreciable, the cosmological component need not be identifiable once astrophysical foregrounds are fitted simultaneously. 
This also demonstrates why the power-law-integrated sensitivity shown in Fig.~\ref{fig:gw-parm-change} cannot by itself be interpreted as a reconstructed threshold.

\begin{figure}[t]
    \centering
    \includegraphics[width=1\linewidth]{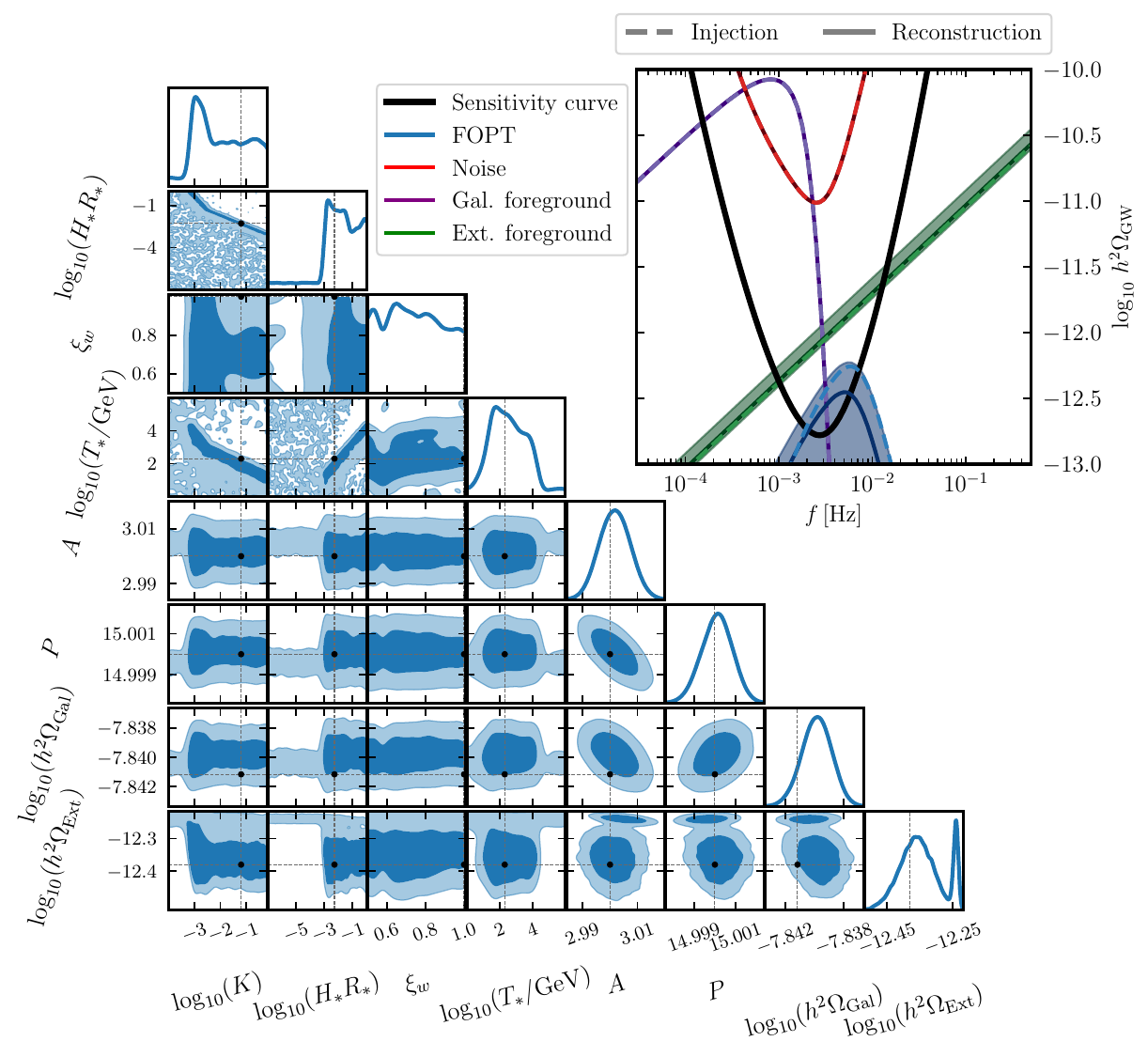}
    \caption{
    Bayesian reconstruction for Case~2 using the broad extragalactic-foreground prior. The corner plot shows the posterior distributions of the \ac{FOPT}, foreground, and instrumental-noise parameters. The upper-right panel shows the reconstructed stochastic components. The upper-right panel shows the reconstructed stochastic components. Under the broad foreground prior, the injected \ac{FOPT} contribution is largely absorbed by the extragalactic compact-binary foreground, resulting in a low reconstructed \ac{SNR} despite the injected value $\rho_{\rm inj} = 19.7$.
    }
    \label{fig:run_2_omega_wide}
\end{figure}

For Case 2, the injected signal has $\rho_{\rm inj} = 19.7$. With the broad extragalactic-foreground prior, however, the reconstructed-SNR posterior is $\rho_{\rm rec} = 16.0^{+7.3}_{-9.4}$, showing that in some cases part of the injected cosmological power is not assigned to the \ac{FOPT} component. 
Figure~\ref{fig:run_2_omega_wide} shows that the part of the \ac{FOPT} spectrum accessible in the most sensitive region of the \ac{LISA} band can be mimicked by the approximately power-law extragalactic compact-binary foreground. 
The two components are consequently difficult to separate when the foreground amplitude is only weakly constrained explaining the large distribution of the recovered SNR.

The loss of information is also apparent in the thermodynamic posteriors,
\begin{align} 
\log_{10}K &= -1.9^{+1.6}_{-1.2}, & \log_{10}(H_*R_*) &= -2.20^{+1.40}_{-0.80}, \\ 
\xi_w &= 0.815^{+0.085}_{-0.076}, & \log_{10}(T_*/{\rm GeV}) &= 2.7^{+1.1}_{-1.1}. 
\end{align}
These broad intervals indicate that little information about the underlying transition parameters is retained once the cosmological component becomes degenerate with the foreground.

\begin{figure}[t]
    \centering
    \includegraphics[width=1\linewidth]{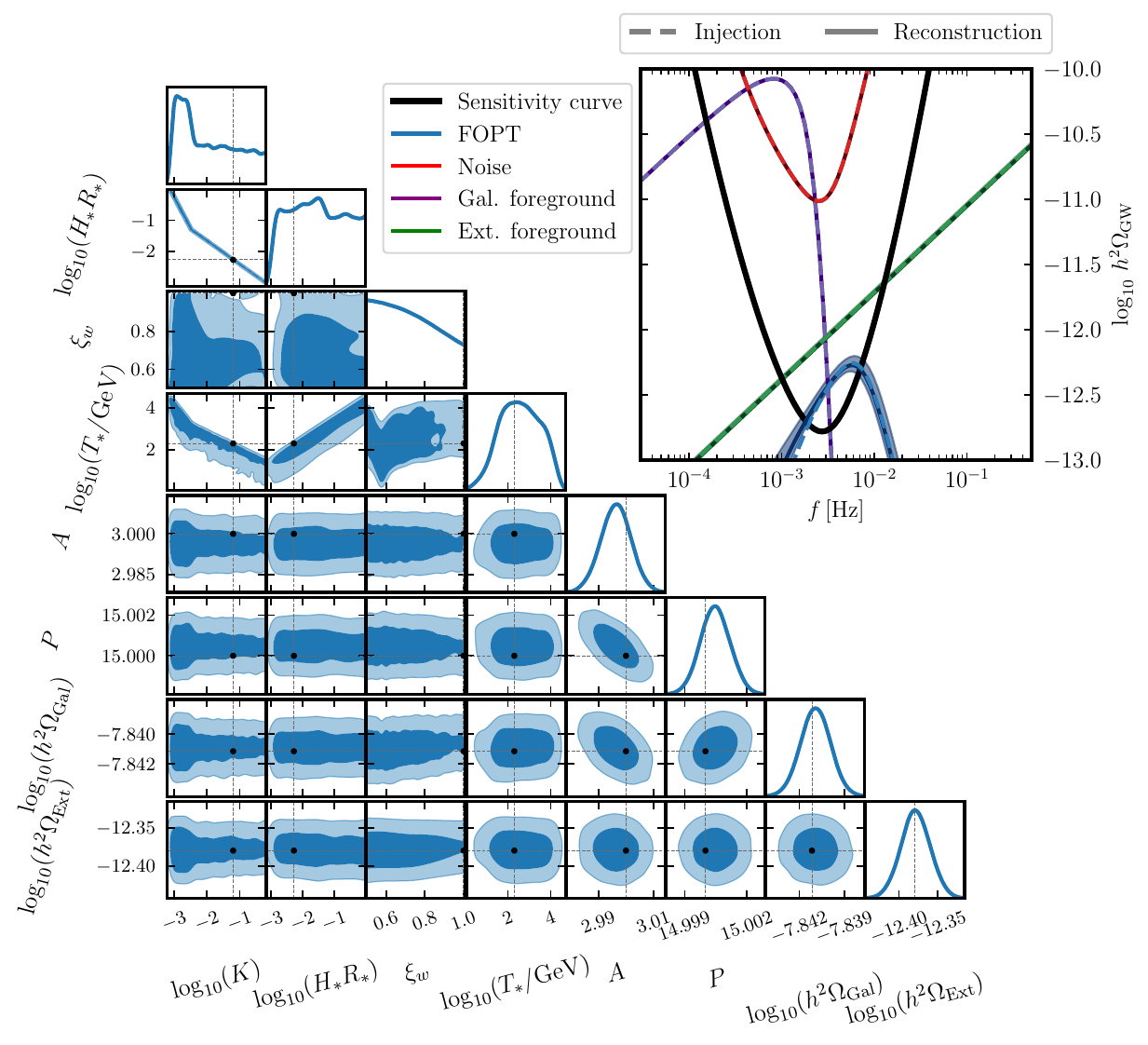}
        \caption{
        Bayesian reconstruction for Case~2 using the narrow extragalactic-foreground prior. The tighter foreground constraint reduces the degeneracy between the compact-binary background and the cosmological component, allowing the \ac{FOPT} spectrum to be reconstructed. The thermodynamic parameters nevertheless remain strongly correlated and only weakly constrained individually.
        }
    \label{fig:run_2_omega_narrow}
\end{figure}

We next repeat the analysis using the informative extragalactic-foreground prior of Section~\ref{sec:foregrounds_noise}. 
For Case~2, this changes the reconstructed-SNR posterior to $\rho_{\rm rec} = 21.9^{+3.3}_{-3.3}$.
The cosmological component can therefore be separated from the foreground once the latter is externally constrained, as illustrated in Fig.~\ref{fig:run_2_omega_narrow}. 

Importantly, the improvement in component separation does not translate into a uniform tightening of all four thermodynamic parameters.
For Case~2, the narrow-prior run gives
\begin{align} 
\log_{10}K &= -2.0^{+1.8}_{-0.62}, & \log_{10}(H_*R_*) &= -1.7^{+1.7}_{-0.48}, \\ 
\xi_w &= 0.84^{+0.16}_{-0.24}, & \log_{10}(T_*/{\rm GeV}) &= 2.8^{+1.4}_{-1.2}. 
\end{align}
The individual thermodynamic posteriors thus remain broad, and some are comparable to or wider than their broad-prior counterparts. The main effect of the informative foreground prior is instead to restore identifiability of the \ac{FOPT} spectral component.

Case~3 provides an even more foreground-dominated example. With $\rho_{\rm inj}=12.8$, the broad-prior analysis does not yield a meaningful reconstruction of the \ac{FOPT} component, and we therefore do not quote a reconstructed \ac{SNR} in Table~\ref{tab:snr_68CL}. At the same time, the inferred extragalactic-foreground amplitude shifts to $\log_{10}(h^2\Omega_{\rm Ext}) = -12.290^{+0.014}_{-0.0025}$, compared with the injected value of approximately $-12.38$, showing that the foreground absorbs the missing cosmological contribution. The thermodynamic posteriors are correspondingly prior dominated.

With the narrow foreground prior, however, the Case~3 signal is recovered with $\rho_{\rm rec} = 11.9^{+3.1}_{-3.0}$, consistent with its injected $\rho_{\rm inj} = 12.8$ within the $68\%$ \ac{CI}. In this case, the informative foreground prior also substantially improves the reconstruction of $K$, $H_* R_*$, and $T_*$, although $\xi_w$ remains weakly constrained. The Case~3 corner plots are shown in Appendix~\ref{sec:third_case}.

Taken together, Cases 2 and 3 demonstrate that the limiting factor for these weaker benchmarks is not simply their optimal SNR, but the ability to distinguish the shape of the cosmological spectrum from that of the extragalactic foreground. External information on the compact-binary background can therefore have a major impact on the reconstruction of \ac{FOPT} signals. At the same time, recovering the cosmological spectral component does not automatically imply precise measurement of the underlying thermodynamic quantities, which remain affected by the intrinsic degeneracies of the sound-wave template.

\begin{table*}[t]
    \centering
    \begin{threeparttable}
    \footnotesize
    \renewcommand{\arraystretch}{1.25}
    \setlength{\tabcolsep}{5pt}
    \begin{tabular}{c cc cccc c}
        \toprule
        & \multicolumn{2}{c}{\ac{SNR}}
        & \multicolumn{4}{c}{\ac{FOPT} parameters}
        & \multicolumn{1}{c}{Foreground} \\
        \cmidrule(lr){2-3}
        \cmidrule(lr){4-7}
        \cmidrule(l){8-8}
        Case
        & $\rho_{\rm inj}$
        & $\rho_{\rm rec}$
        & $\log_{10}(K)$
        & $\log_{10}(H_*R_*)$
        & $\xi_w$
        & $\log_{10}(T_*/\mathrm{GeV})$
        & $\log_{10}(h^2\Omega_{\rm Ext})$ \\
        \midrule

        1
        & $32.3$
        & $35.8^{+6.9}_{-7.1}$
        & $-1.6^{+1.3}_{-1.1}$
        & $-1.88^{+0.59}_{-0.92}$
        & $0.792^{+0.180}_{-0.087}$
        & $2.44^{+0.61}_{-0.84}$
        & $-12.383^{+0.054}_{-0.045}$ \\
        \addlinespace

        2
        & \multirow{2}{*}{$19.7$}
        & $16.0^{+7.3}_{-9.4}$
        & $-1.9^{+1.6}_{-1.2}$
        & $-2.20^{+1.40}_{-0.80}$
        & $0.815^{+0.085}_{-0.076}$
        & $2.7^{+1.1}_{-1.1}$
        & $-12.360^{+0.130}_{-0.094}$ \\

        2 (NP)
        &
        & $21.9^{+3.3}_{-3.3}$
        & $-2.0^{+1.8}_{-0.62}$
        & $-1.7^{+1.7}_{-0.48}$
        & $0.84^{+0.16}_{-0.24}$
        & $2.8^{+1.4}_{-1.2}$
        & $-12.385^{+0.017}_{-0.017}$ \\
        \addlinespace

        3
        & \multirow{2}{*}{$12.8$}
        & \textemdash
        & $-2.3^{+2.1}_{-1.7}$
        & $-3.7^{+3.7}_{-1.3}$
        & $0.75^{+0.25}_{-0.15}$
        & $3.0^{+3.0}_{-3.0}$
        & $-12.290^{+0.014}_{-0.003}$ \\

        3 (NP)
        &
        & $11.9^{+3.1}_{-3.0}$
        & $-1.74^{+0.86}_{-0.86}$
        & $-1.87^{+0.52}_{-1.10}$
        & $0.76^{+0.24}_{-0.16}$
        & $2.42^{+0.73}_{-0.99}$
        & $-12.376^{+0.019}_{-0.019}$ \\

        \bottomrule
    \end{tabular}

    \begin{tablenotes}[flushleft]
        \footnotesize
        \item Posterior medians and corresponding $68\%$ \ac{CI} for the \ac{FOPT} parameters and extragalactic-foreground amplitude. The label NP denotes the analyses employing the narrow extragalactic-foreground prior introduced in Section~\ref{sec:foregrounds_noise}. The instrumental-noise and galactic-foreground parameters are omitted because they are tightly constrained in all runs. Their relative uncertainty at $68\%$ \ac{CI} is at most 0.3\%, $0.007\%$, and $ 0.03\%$ for $A$, $P$, and $\log_{10}(h^2\Omega_{\rm Gal})$, respectively . 
        The values of $\rho_{\rm rec}$ are the median and $68\%$ \ac{CI} of the derived posterior distribution $p(\rho_{\rm rec}\mid D)$, obtained by evaluating Eq.~\eqref{eq:SNR_rec} for each posterior sample. No reconstructed-SNR value is quoted for Case~3 with the broad foreground prior because the \ac{FOPT} component is not meaningfully reconstructed.
    \end{tablenotes}

    \caption{
    Reconstructed \ac{FOPT} and extragalactic-foreground parameters for the three benchmark cases. The injected \ac{SNR} is also reported for comparison.
    }
    \label{tab:snr_68CL}
    \end{threeparttable}
\end{table*}

\section{Conclusions}
\label{conclusions}

In this work, we investigated the reconstruction with \ac{LISA} of stochastic \ac{GW} signals generated by \acp{FOPT} in a \ac{FD} model. 
Starting from the finite-temperature effective potential, we identified representative TeV-scale phase transitions and mapped their thermodynamic parameters onto the corresponding sound-wave \ac{SGWB} spectra. 
We then injected three benchmark signals into simulated \ac{LISA} data and performed a joint Bayesian reconstruction of the cosmological signal, astrophysical foregrounds, and instrumental noise using \texttt{SGWBinner}.

The strongest benchmark, with injected $\rho_{\rm inj}=32.3$, can be successfully reconstructed in the presence of the assumed foregrounds and instrumental noise. 
Its reconstructed-SNR posterior is $\rho_{\rm rec} = 35.8^{+6.9}_{-7.1}$. The individual thermodynamic parameters nevertheless remain strongly correlated and only weakly constrained. 
This reflects an intrinsic property of the sound-wave template: four thermodynamic quantities, $\{K,H_*R_*,\xi_w,T_*\}$, determine only three independent spectral parameters. 
Quantities that characterize the observable spectrum directly are consequently better constrained than the underlying thermodynamic parameters. 
A measurement of the stochastic spectrum would therefore not, by itself, imply a comparable precise determination of the microscopic properties of the transition.

For the weaker benchmarks, the dominant limitation is the degeneracy with the unresolved extragalactic compact-binary foreground. 
With the broad foreground prior, Case~2 has an injected $\rho_{\rm inj} = 19.7$, but yields $\rho_{\rm rec} = 16.0^{+7.3}_{-9.4}$, while for Case~3 ($\rho_{\rm inj} = 12.8$) the cosmological component is not meaningfully reconstructed.
These examples also demonstrate that lying above  a power-law-integrated sensitivity curve does not guarantee that a cosmological component can be separated from astrophysical foregrounds in a multi-component analysis.

External information on the extragalactic foreground can substantially change this conclusion. 
With the informative foreground prior, the reconstructed SNR becomes $\rho_{\rm rec} = 21.9^{+3.3}_{-3.3}$ for Case~2 and $\rho_{\rm rec} = 11.9^{+3.1}_{-3.0}$ for Case~3. The cosmological component can therefore be recovered in both cases. For Case~2, however, the individual thermodynamic parameters remain broad despite the improved spectral separation, whereas Case~3 also shows a substantial tightening of several thermodynamic posteriors. This distinction emphasized that separating a primordial component from the foreground and inferring the physical properties of the underlying phase transition are related but distinct inference problems.

Future work can extend this analysis along several complementary directions. 
On the theoretical side, a dedicated calculation of the bubble-wall velocity would allow the relativistic-wall assumption adopted here to be relaxed \cite{DeCurtis:2022hlx, Laurent:2022jrs, Ai:2023see, DeCurtis:2024hvh, Krajewski:2024gma,Ai:2025bjw, Carena:2025flp}, while including a possible turbulent contribution and higher-order thermal corrections would provide a more complete modelling of both the \ac{PT} dynamics and the resulting \ac{GW} spectrum \cite{Caprini:2026nnk, Stomberg:2025kxf, Niksa:2018ofa, Parwani:1991gq, Arnold:1992rz, Wainwright:2011qy, Correia:2025qif}. 
On the data-analysis side, it will be important to assess the robustness of the reconstruction against mismatches between the injected and recovered signal, foreground, and instrumental-noise models \cite{Kume:Assessing_impact_2025}, as well as to include residuals from the subtraction of individually resolved sources \cite{Rosati:SGWB_recovery_2024}. 
Concerning the treatment of the astrophysical foregrounds, the inclusion of the annual modulation in the Galactic compact-binary foreground has been shown to substantially improve the reconstruction of the cosmological signal~\cite{Hindmarsh:2024ttn}.
There are also additional astrophysical backgrounds that should be taken into account by future studies, such as the EMRI background and the background of unresolved massive black hole binaries~\cite{Sesana:2004sp,Sesana:2004gf,Bonetti:2020jku,Pozzoli:2023kxy,Piarulli:2024yhj}.
These developments would enable a more realistic quantification of the theoretical and observational uncertainties affecting the reconstruction of \ac{FD} models.

A further natural extension is to complement parameter reconstruction with an explicit Bayesian model comparison between hypotheses with and without a cosmological \ac{FOPT} component. This would allow the reconstructed-\ac{SNR} diagnostic used here to be supplemented by a formal assessment of the statistical preference for a primordial signal. 
Combining such a model-selection analysis with improved astrophysical foreground models and increasingly informative external constraints on the compact-binary background would provide a more complete framework for assessing the detectability of \ac{FOPT} signals with \ac{LISA}.

Overall, our results show that TeV-scale \ac{FD} scenarios can produce stochastic \ac{GW} spectra that are reconstructable with \ac{LISA}, while also demonstrating that their observability can be limited as much by component degeneracies as by the intrinsic signal amplitude. Accurate modelling and independent constraints on the astrophysical foreground will therefore be central to extracting particle-physics information from a future cosmological stochastic background.

\acknowledgments

The authors are grateful to Mauro Pieroni for useful comments and discussions and for the assistance with \texttt{SGWBinner}. NF is grateful to Chiara Caprini and Alberto Roper Pol for useful discussions. DL acknowledges funding from the UZH Postdoc Grant no. [K-72341-01-01].

\appendix
\section{Estimate of the turbulent contribution}
\label{sec:turbulent_motion}

In the baseline analysis we approximate the \ac{FOPT} signal by its sound-wave contribution,
\begin{equation}
    \Omega_{\rm FOPT}(f) \simeq \Omega_{\rm sw}(f),
\end{equation}
and neglect the stochastic background generated by \ac{MHD} turbulence. In this Appendix, we estimate the size of the turbulent component for the benchmark transitions cosidered in this work.

We denote by $\epsilon$ the fraction of the bulk kinetic energy that is transferred to vortical motions and subsequently sources \ac{MHD} turbulence.
Following the estimates commonly adopted in the literature~\cite{Caprini:2015zlo, Hindmarsh:2015qta, Hindmarsh:2017gnf, Hindmarsh:2013xza}, we take
$\epsilon = 0.1$, which provides a conservative benchmark for the present estimate.

For the short-lived acoustic regime relevant to our benchmarks, Eq.~\eqref{eq:sw_amplitude_int} gives the logarithmically integrated sound-wave amplitude,
\begin{equation} 
h^2\Omega_{\rm sw} = h^2 A_{\rm sw} \left(\frac{a_*}{a_0}\right)^4 \left(\frac{H_*}{H_0}\right)^2 \frac{K^2(H_*R_*)^2}{\bar v_f}, 
\end{equation}
where expressions for $\bar v_f$ and $\Gamma$ are given in Eq.~\eqref{eq:sound_lifetime}.
For the double-broken-power-law template adopted in this work, the sound-wave amplitude evaluated at the second break is approximately related to the integrated amplitude by~\cite{Caprini:2024hue}
\begin{equation} 
h^2\Omega_2 \simeq 0.55\,h^2\Omega_{\rm sw}, 
\end{equation} 
where $\Omega_2\equiv\Omega_{\rm sw}(f_2)$, as defined in Eq.~\eqref{eq:omega_dbpl_reparam}.

Within the turbulent template of Ref.~\cite{Caprini:2024hue}, the peak amplitude of the \ac{MHD} contribution is 
\begin{equation} 
h^2\Omega_{\rm turb,peak} = h^2 A_{\rm turb} \left(\frac{a_*}{a_0}\right)^4 \left(\frac{H_*}{H_0}\right)^2 (\epsilon K)^2 (H_*R_*)^2, 
\end{equation} 
with 
\begin{equation} A_{\rm turb}\simeq4.37\times10^{-3}. 
\end{equation}
Dividing this expression by the sound-wave amplitude at the second break gives
\begin{equation} 
\label{eq:turb_sound_ratio} 
\frac{\Omega_{\rm turb,peak}}{\Omega_2} = \frac{A_{\rm turb}} {0.55\,A_{\rm sw}} \epsilon^2\bar v_f. 
\end{equation}

The ratio in Eq.~\eqref{eq:turb_sound_ratio} increases with the characteristic fluid velocity and is therefore largest for the strongest transition among our benchmark cases. For Case~1, $\alpha=0.323$. Using Eqs.~\eqref{eq:K} and~\eqref{eq:kappa}, we obtain 
\begin{equation} K\simeq7.2\times10^{-2}, \qquad \bar v_f = \sqrt{\frac{3K}{4}} \simeq0.23. 
\end{equation}
Taking $A_{\rm sw}=0.11$, $A_{\rm turb}=4.37\times10^{-3}$, and $\epsilon=0.1$, Eq.~\eqref{eq:turb_sound_ratio} then gives 
\begin{equation} 
\frac{\Omega_{\rm turb,peak}}{\Omega_2} \simeq 1.7\times10^{-4}. 
\end{equation}

Within the modelling prescription adopted here, the turbulent contribution is therefore approximately four orders of magnitude smaller than the sound-wave amplitude at the second spectral break, even for the strongest benchmark transition. 
This estimate supports the sound-wave-only approximation used in the baseline analysis. 
Nevertheless, because the benchmark transitions lie in the short-lived acoustic regime, the treatment of the subsequent conversion of acoustic motion into vorticity remains a modelling uncertainty. A more complete treatment of the turbulent contribution is therefore a natural extension of the present analysis.

\section{Power-law sensitivity curve} \label{app:pli}

The \ac{PLS} provides a graphical representation of the sensitivity to stochastic backgrounds whose spectra can be approximated by power laws over the detector band \citep{Thrane:Sensitivty_curves_2013, Caprini:Reconstructing_spectral_shape_2019}. 
Unlike a pointwise noise curve, it accounts for the increase in signal-to-noise ratio obtained by integrating over frequency and observation time. We consider a family of power-law backgrounds, 
\begin{equation} 
\label{eq:pli_power_law} 
h^2\Omega_{\beta}(f) = h^2\Omega_{\beta} \left( \frac{f}{f_{\rm ref}} \right)^{\beta}, 
\end{equation} 
where $\beta$ is the spectral index, $f_{\rm ref}$ is an arbitrary reference frequency, and $\Omega_{\beta}$ denotes the amplitude at $f_{\rm ref}$. 
We use $f_{\rm ref}=1~\mathrm{mHz}$ for numerical convenience. The final power-law-integrated curve is independent of this choice, provided that the amplitude is consistently redefined. Using the same convention as for the optimal SNR in Eq.~\eqref{eq:SNR}, the SNR of the power law in Eq.~\eqref{eq:pli_power_law} is 
\begin{equation} 
\label{eq:pli_snr} 
\rho_{\beta}^2 = T_{\rm obs} \sum_{I\in\mathcal{I}} \int_{f_{\min}}^{f_{\max}} df\, \left[ \frac{h^2 \Omega_{\beta} (f/f_{\rm ref})^{\beta} }{ h^2\Omega_{N,I}(f) } \right]^2. 
\end{equation} 
Here, $\mathcal{I}$ denotes the set of TDI channels included in the calculation and $h^2\Omega_{N,I}(f)$ is the equivalent fractional-energy-density noise spectrum in channel $I$, including the appropriate \ac{LISA} response function. In the present analysis, we use the same channels, frequency interval, instrumental-noise model, and observation time as in Eq.~\eqref{eq:SNR}. 
For each spectral index $\beta$, the amplitude is chosen such that the power-law signal reaches a prescribed threshold $\rho_{\rm th}$. Solving Eq.~\eqref{eq:pli_snr} for this amplitude gives 
\begin{equation} 
\label{eq:pli_amplitude} 
\Omega_{\beta}^{\rm th} = \rho_{\rm th} \left\{ T_{\rm obs} \sum_{I\in\mathcal{I}} \int_{f_{\min}}^{f_{\max}} df\, \left[ \frac{ (f/f_{\rm ref})^{\beta} }{ \Omega_{N,I}(f) } \right]^2 \right\}^{-1/2}. 
\end{equation} 
The corresponding threshold power law is 
\begin{equation} 
\label{eq:pli_threshold_power_law} h^2\Omega_{\beta}^{\rm th}(f) = h^2\Omega_{\beta}^{\rm th} \left( \frac{f}{f_{\rm ref}} \right)^{\beta}. 
\end{equation} 
The \ac{PLS} is obtained by taking the upper envelope of the threshold power laws at each frequency, 
\begin{equation} 
\label{eq:pli_envelope} h^2\Omega_{\rm PLS}(f) = \max_{\beta} \left[ h^2\Omega_{\beta}^{\rm th}(f) \right]. 
\end{equation} 
In practice, Eq.~\eqref{eq:pli_amplitude} is evaluated on a dense grid of spectral indices covering the range $\beta_{\min}\leq\beta\leq\beta_{\max}$ used in the numerical implementation. 
For the blue curve in Fig.~\ref{fig:gw-parm-change}, we adopt \begin{equation} T_{\rm obs}=4~\mathrm{yr}, \qquad \rho_{\rm th}=10, \end{equation} and use instrumental noise only in the denominator of Eq.~\eqref{eq:pli_snr}. 
A power-law background tangent to the resulting curve has $\rho=\rho_{\rm th}$, while a power law that intersects the region above the curve has $\rho>\rho_{\rm th}$ under the assumptions entering its construction. The \ac{FOPT} spectra analysed in this work are double broken power laws rather than single power laws. Their comparison with $h^2\Omega_{\rm PLS}(f)$ therefore provides only a qualitative indication of their broadband strength. Their actual optimal SNR must be evaluated using the full spectrum in Eq.~\eqref{eq:SNR}.

\section{Fisher-matrix approximation} \label{app:fisher}

We complement the Bayesian analysis with a Fisher-information forecast, following the Gaussian approximation used in Ref.~\cite{Caprini:2024hue, Gowling:Fisher_prospects_2021, Boileau:Prospects_FOPT_detection_2022, Gowling:Reconstructing_physical_parameters_2022}. The Fisher approach provides an estimate of the local curvature of the likelihood around a fiducial parameter vector $\boldsymbol{\theta}_{\rm fid}$ and is expected to be reliable only when the posterior is approximately Gaussian and sufficiently far from prior boundaries. Defining $\Delta\boldsymbol{\theta} = \boldsymbol{\theta} - \boldsymbol{\theta}_{\rm fid}$ and expanding the log-likelihood to second order around the fiducial point gives 
\begin{equation} 
\label{eq:fisher_likelihood_expansion} 
-2\ln\mathcal{L}(\boldsymbol{\theta}) \simeq -2\ln\mathcal{L}(\boldsymbol{\theta}_{\rm fid}) + \sum_{a,b} F_{ab}\, \Delta\theta_a\Delta\theta_b, 
\end{equation} 
where $F_{ab}$ is the Fisher information matrix. For each TDI channel $I$, we denote the total model in fractional-energy-density units by 
\begin{equation} 
\label{eq:fisher_total_spectrum} 
\mathcal{D}_I(f;\boldsymbol{\theta}) ={} h^2\Omega_{{\rm FOPT},I}(f;\boldsymbol{\theta}) + h^2\Omega_{{\rm Ext},I}(f;\boldsymbol{\theta})  + h^2\Omega_{{\rm Gal},I}(f;\boldsymbol{\theta}) + h^2\Omega_{N,I}(f;\boldsymbol{\theta}).  
\end{equation} 
The response functions appropriate to each channel are understood to be included in the individual terms. In the continuous-frequency Gaussian approximation, the data contribution to the Fisher matrix is 
\begin{equation} 
\label{eq:fisher_matrix} 
F_{ab}^{\rm data} = T_{\rm obs} \sum_{I\in\mathcal{I}} \int_{f_{\min}}^{f_{\max}} df\, \frac{ \partial_a\mathcal{D}_I(f) \, \partial_b\mathcal{D}_I(f) }{ \mathcal{D}_I^2(f) } \bigg|_{\boldsymbol{\theta} =\boldsymbol{\theta}_{\rm fid}}, 
\end{equation} 
where $\partial_a \equiv \partial/{\partial\theta_a}$. 
The parameter vector contains the \ac{FOPT}, foreground, and instrumental-noise parameters varied in the corresponding Bayesian run. For example, for the geometric reconstruction it is 
\begin{equation} 
\label{eq:fisher_parameter_vector} 
\boldsymbol{\theta} =\bigl\{\log_{10}(\Omega_2),\, \log_{10}(f_1),\, \log_{10}(f_2), \log_{10}(h^2\Omega_{\rm Ext}),\, \log_{10}(h^2\Omega_{\rm Gal}),\, A,\, P \bigr\}. 
\end{equation} 
When Gaussian prior information is included, its curvature can be added directly to the data Fisher matrix, 
$F_{ab} = F_{ab}^{\rm data} + F_{ab}^{\rm prior}$. 
For independent Gaussian priors with standard deviations $\sigma_{a,\rm prior}$, 
\begin{equation} 
\label{eq:fisher_prior_matrix} 
F_{ab}^{\rm prior} = \frac{\delta_{ab}} {\sigma_{a,\rm prior}^2}. 
\end{equation} 
Equation~\eqref{eq:fisher_prior_matrix} is applied to the galactic and extragalactic foreground amplitudes and to the instrumental-noise parameters using the prior widths in Table~\ref{tab:foreground_noise_priors}. 
The covariance matrix predicted by the Fisher approximation is $C \simeq F^{-1}$. 
The marginalized one-dimensional standard deviation of parameter $\theta_a$ is therefore 
\begin{equation} 
\sigma(\theta_a) = \sqrt{C_{aa}}. 
\end{equation} 
For a parameter sampled in base-10 logarithmic form, $x_a=\log_{10}q_a$, the corresponding fractional uncertainty is, in the small-error limit, 
\begin{equation} 
\frac{\sigma(q_a)}{q_a} \simeq \ln(10)\, \sigma(x_a). 
\end{equation} 
Marginalized two-dimensional Fisher contours for parameters $(\theta_a,\theta_b)$ as shown in Figure \ref{fig:relative_error_fisher}  are obtained from the corresponding $2\times2$ sub-block of the full covariance matrix.

\section{Reconstruction of Case 3}
\label{sec:third_case}

\begin{figure}[t]
    \centering
    \includegraphics[width=1.\linewidth]{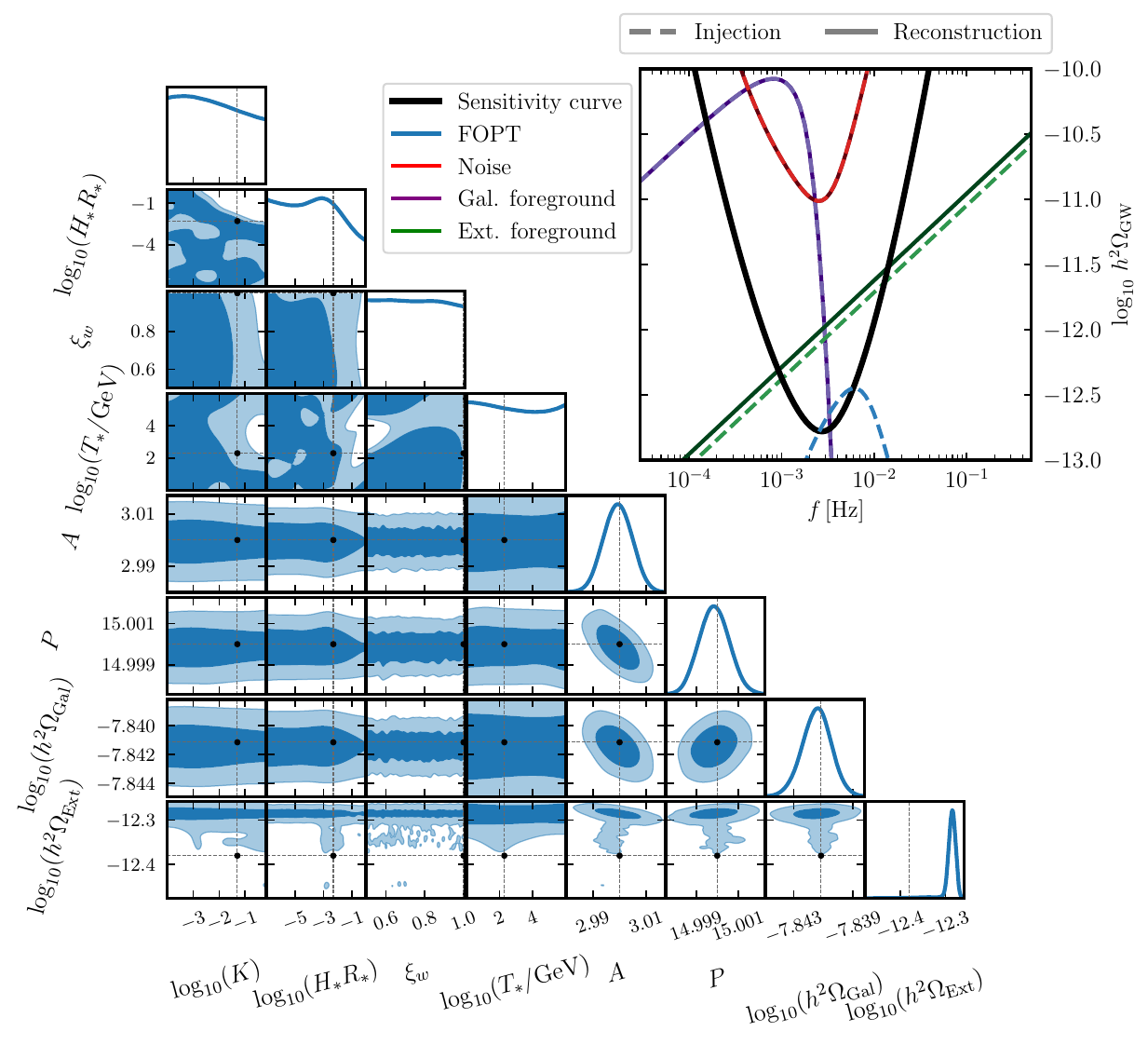}
    \caption{
    Bayesian reconstruction for Case~3 using the broad extragalactic-foreground prior. The corner plot shows the posterior distributions of the \ac{FOPT}, astrophysical foreground, and instrumental-noise parameters. In the upper-right spectral reconstruction, the cosmological component is not meaningfully recovered and is largely under-evaluated, its contribution being absorbed by the extragalactic compact-binary foreground.
    }
    \label{fig:third_run_wide_prior}
\end{figure}

\begin{figure}[t]
    \centering
    \includegraphics[width=1.\linewidth]{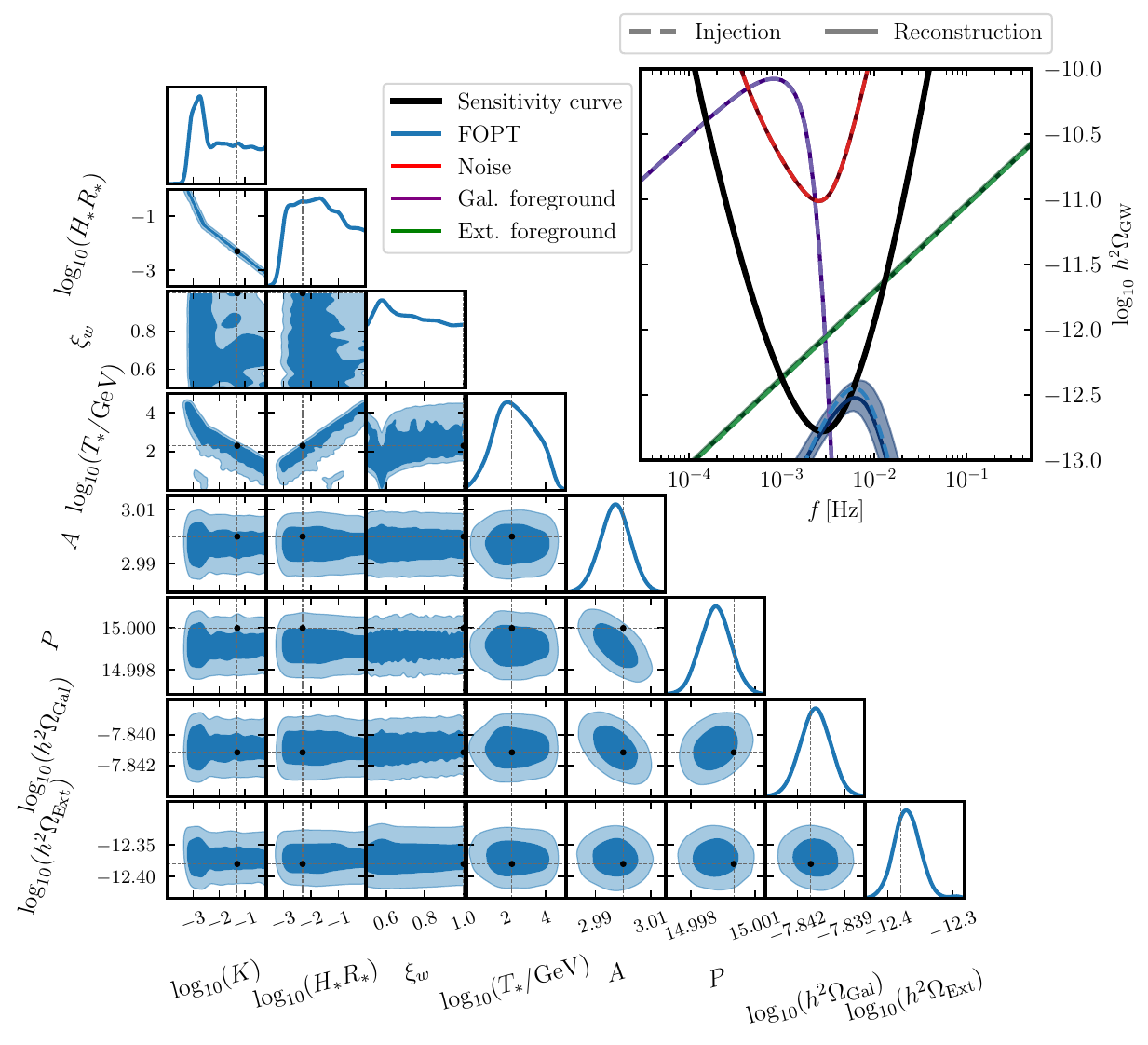}
    \caption{
    Bayesian reconstruction for Case~3 using the narrow extragalactic-foreground prior. The tighter foreground constraint breaks the dominant degeneracy with the cosmological component, allowing the injected \ac{FOPT} spectrum to be reconstructed. The resulting reconstructed-SNR posterior is $\rho_{\rm rec} = 11.9^{+3.1}_{-3.0}$, compared with $\rho_{\rm inj}=12.8$. 
    }
    \label{fig:third_run_narrow_prior}
\end{figure}

Here we provide the full posterior distributions for Case~3, the weakest of the three benchmark signals considered in the main analysis. Its injected \ac{SNR} is $\rho_{\rm inj} = 12.8$. Figures~\ref{fig:third_run_wide_prior} and~\ref{fig:third_run_narrow_prior} show the reconstruction using, respectively, the broad and narrow priors on the extragalactic compact-binary foreground.

With the broad prior, the \ac{FOPT} component cannot be meaningfully separated from the extragalactic foreground. As shown in Fig.~\ref{fig:third_run_wide_prior}, the foreground amplitude shifts away from its injected value and absorbs the additional stochastic power. Correspondingly, the signal is under-evaluated and the thermodynamic parameters are almost entirely prior dominated; see Table~\ref{tab:snr_68CL}. We therefore do not quote a reconstructed-SNR posterior for this run.

The situation changes when the informative extragalactic-foreground prior of Section~\ref{sec:foregrounds_noise} is imposed. Figure~\ref{fig:third_run_narrow_prior} shows that the cosmological component can then be separated from the foreground. The resulting reconstructed-SNR posterior is $\rho_{\rm rec} = 11.9^{+3.1}_{-3.0}$, consistent with the injected value within the $68\%$ \ac{CI}. The posteriors of $K$, $H_*R_*$, and $T_*$ also tighten substantially relative to the broad-prior analysis, whereas the bubble-wall velocity remains weakly constrained.

\newpage

\bibliographystyle{JHEP}
\bibliography{bib}

\end{document}